\documentclass[conference]{IEEEtran}

\ifCLASSINFOpdf
\else
\fi
\usepackage{url} 
\usepackage{algorithm}
\usepackage{algorithmic}

\usepackage{multirow} 
\usepackage{adjustbox}

\usepackage{subcaption}
 \usepackage{booktabs}
\usepackage{hyperref}

\hypersetup{
    colorlinks=true,
    citecolor=red,
    linkcolor=green,
    urlcolor=black
}

\usepackage{makecell} 

\begin{document}
%
\title{Code-Poisoning Property Inference Attacks}

	

%
\author{\IEEEauthorblockN{Xukun Luan\IEEEauthorrefmark{2},
Yuhui Gong\IEEEauthorrefmark{2},
Gang Zhang\IEEEauthorrefmark{2},
Zixuan Huang\IEEEauthorrefmark{4},
Yuanguo Bi\IEEEauthorrefmark{4},
Xuesong Li\IEEEauthorrefmark{2} and
Jinyan Liu\IEEEauthorrefmark{2}}
\IEEEauthorblockA{\IEEEauthorrefmark{2}School of Computer Science and Technology, Beijing Institute of Technology}
\IEEEauthorblockA{\IEEEauthorrefmark{4}School of Computer Science and Engineering, Northeastern University}
Email: xukunluan@bit.edu.cn
}



\maketitle


\begin{abstract}
The flourishing code hosting platforms and coding agents enable even beginners with private data to build tailored Machine Learning (ML) models using available code quickly. The training data for ML models, often regarded as private property (e.g., clinical records, transaction information), is at significant risk of information leakage. Property Inference Attacks (PIAs), as a significant type of privacy attack, aim to expose global property information of the training set. In this paper, we present Code‑Poisoning Property Inference Attack (CPPIA), the first code‑level PIA, which overcomes four limitations of existing works: insufficient attack performance, severe degradation of model accuracy, high computational overhead, and failure under defenses. We consider malicious code providers from code-hosting platforms (GitHub) and coding agents (Codex). Upon downloading the poisoned code, data holders train models with their private data without professional auditing, subsequently releasing label-only APIs to the public. The adversary embeds the properties into secret samples during training and queries the trained model on these samples later to leak privacy. CPPIA offers 100\% attack accuracy without degrading model accuracy. It is also computationally lightweight and requires no shadow models. We evaluate the attack performance across four datasets, eight model architectures, eighteen properties, and under three defense mechanisms, demonstrating the universality and effectiveness of CPPIA\footnote{Our code is available at \url{https://github.com/Zili1000/CPPIA}.}.
\end{abstract}



%
\IEEEpeerreviewmaketitle

\section{Introduction}
The advent of code hosting platforms and coding agents has offered programmers unprecedented convenience while significantly reducing the barrier to entry for beginners. In practice, developing Machine Learning (ML)~\cite{jordan2015machine} models from scratch is uncommon; it has become standard practice to utilize tools such as GitHub~\cite{github} and Codex~\cite{Codex} in the development pipeline. Nevertheless, the very convenience afforded by these platforms and agents comes with inherent security vulnerabilities~\cite{wwww,eeee}. Misconfigurations can be exploited by an adversary to execute a supply chain attack against TensorFlow releases on both GitHub and PyPi by compromising TensorFlow's build agents through a malicious pull request~\cite{ttttt}. Furthermore, Anthropic has incorporated undisclosed detection mechanisms into its Claude Code~\cite{Claude} CLI tool, which are specifically designed to target Chinese users or traffic routed through Chinese AI laboratory proxies~\cite{ClaudePPPPPP}. In one incident, the adversary compromises a Linux server and repurposes it as a transient host, running local instances of Claude and Codex, ultimately recovering the complete agent directory, associated tooling, and more than one thousand session logs~\cite{PPPPPPPPPPPP}. The aforementioned code-level vulnerabilities have raised significant concerns about potential privacy compromises throughout the lifecycle of ML model development.

In the development of ML~\cite{jordan2015machine}, the abundance of code hosting platforms and coding agents offers great convenience to data holders. At the same time, the sensitive information of data holders for ML models may face unnecessary privacy leakage risks. A common and highly threatening type of attack is Property Inference Attacks (PIAs)~\cite{ganju2018property,zhang2021leakage,mahloujifar2022property}, where an adversary aims to infer global properties of the training set of a target model, including inferring the proportion of rare patients within a certain population or the ratio of individuals with a specific religious affiliation. By leaking such dataset-level properties, adversaries may obtain exploitable sensitive information. For instance, a political party can conduct a PIA on an ML model released by an agency in a swing state, thereby inferring the political inclination distribution within that state, and subsequently adjust its campaign strategy in a targeted manner to gain an advantage.


Some PIAs~\cite{ganju2018property,snap,tian2023manipulating} assume that adversaries typically have access to the model’s internal parameters, output logits, or training set; importantly, executing such attacks requires substantial computational resources, and their performance remains far from satisfactory. Four limitations remain unaddressed: suboptimal attack performance, degradation of model accuracy, high computational cost, and vulnerability to defenses. In this work, we propose a novel code-level PIA, CPPIA, based on code poisoning, which is a new attack vector introduced here for the first time to address the above limitations, as shown in Figure \ref{fig:Overview}. Code poisoning attacks~\cite{bagdasaryan2021blind,song2017machine,11029807} have been extensively studied. A user survey presented in~\cite{liu2022loneneuron} shows that in the vast majority of cases, ML data holders adopt third-party code “as-is,” primarily checking the trained model’s performance on their own domain-specific datasets without scrutinizing the code itself. More importantly, modern ML codebases grow increasingly complex. Gu \textit{et al.}~\cite{gu2026spacezonedemystifyingsecurity} discover that thousands of applications on the platforms are developed by untrustworthy third parties, hundreds contain input injection vulnerabilities that permit arbitrary code execution, and dozens have been found to actively embed backdoors. Coding agents mandate user scrutiny and authorization at each code modification step. Nevertheless, when inundated with a multitude of incremental prompts, users exhibit a strong propensity to approve all requested actions en masse—a behavioral tendency that is frequently attested in practice~\cite{qu2026overeagercodingagentsmeasuring,huang2026codereuseinvestigatingcode}. Poisoned code injected into large, poorly readable repositories can be difficult to detect, and real-world incidents~\cite{wwww,eeee,ttttt} continue to emerge, highlighting the non-negligible research significance of code poisoning attacks.

Analyzing the characteristics of PIAs helps us extend code-poisoning. PIAs target global properties of the training set. Therefore, only a small number of synthetic samples is required to achieve the attack goal (merely 10 synthetic samples suffice to determine the target property's proportion with precision up to three decimal places, regardless of the training set size, as shown in Section \ref{Methodology}). Prior works~\cite{mahloujifar2022property,snap} attempt to enhance data-level PIA performance through data poisoning. The common idea is to craft a set of poisoned samples and inject them into the training set, forcing the model's output to reveal more information about the target property (e.g., increasing the divergence between the output logits of models trained with and without the target property). Tian \textit{et al.}~\cite{tian2023manipulating} employed model poisoning to manipulate the upstream model, thereby achieving model-level PIAs. Our proposed attack aims to address the following four limitations of aforementioned works: (1) the performance of property inference remains insufficient; (2) data poisoning and model poisoning often lead to a significant and potentially unacceptable drop in model performance, which may raise suspicion; (3) the computational cost required for the attack is prohibitively high, hindering practical feasibility; and (4) existing defense mechanisms can easily defend against these attacks. Our contributions are summarized as follows:

\begin{figure}[t]
    \centering
    \includegraphics[width=0.45\textwidth]{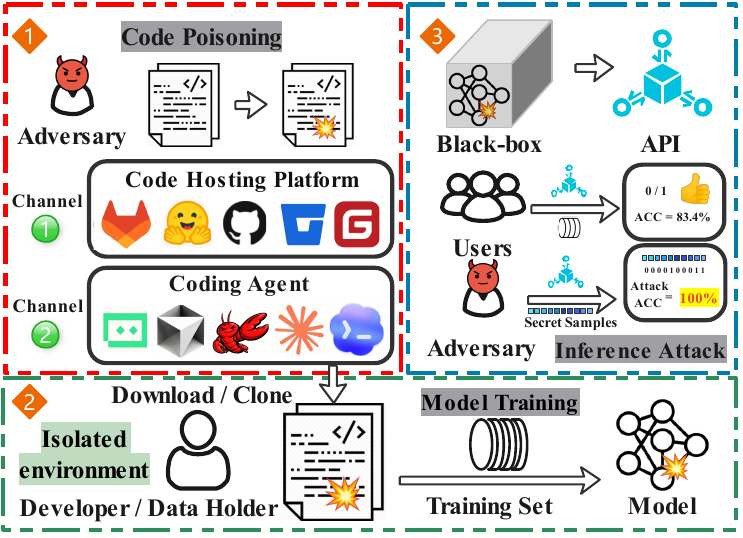}
    \caption{Overview of CPPIA. \textbf{Channel} 1: The adversary publishes a repository containing the poisoned code on public code hosting platforms. \textbf{Channel} 2: Coding agents actively inject poisoned code or are hijacked by hackers to passively introduce poisoned code. A data holder, acting as the developer, downloads and executes this code to train their model. The code-poisoned model can achieve lossless accuracy for users and 100\% PIAs accuracy for the adversary.}
    \label{fig:Overview}
\end{figure}


\begin{itemize}
    \item We execute a nearly perfect property inference attack for the first time, which overcomes the four limitations of the existing works. The adversary can achieve 100\% attack accuracy without degrading the model's performance. The inference attack can be launched with minimal computational overhead and exhibits robustness against defenses.
    \item We propose the first code-level property inference attack, CPPIA. We assume the code is executed in a secure, isolated environment. The adversary cannot directly access this environment. The only outcome of this process is the trained model, to which the adversary has only black-box, label-only query access.
    \item We comprehensively evaluate our attack under a wide range of settings, including 4 datasets, 8 model architectures, 18 different property types, and 3 categories of defense methods. CPPIA is the first to demonstrate that a capable adversary can achieve maximal leakage of property information, thereby exposing significant vulnerabilities in existing property privacy protections.
    \item We demonstrate that the CPPIA can be generalized across a variety of tasks, including image generation, text-to-image, regression, and natural language processing. The implications of CPPIA are far-reaching, and we urge the ML community to prioritize code auditing as a critical security practice.
\end{itemize}

\noindent \textbf{Disclaimer}: \textit{This paper is provided for scholarly reference purposes only. We disclaim any liability for potential privacy breaches arising from the use of code hosting platforms or coding agents in real-world deployments.}

\section{Preliminaries and Related Work}

\subsection{Code Hosting Platforms and Coding Agents}

\begin{figure}[t]
    \centering
    \includegraphics[width=0.45\textwidth]{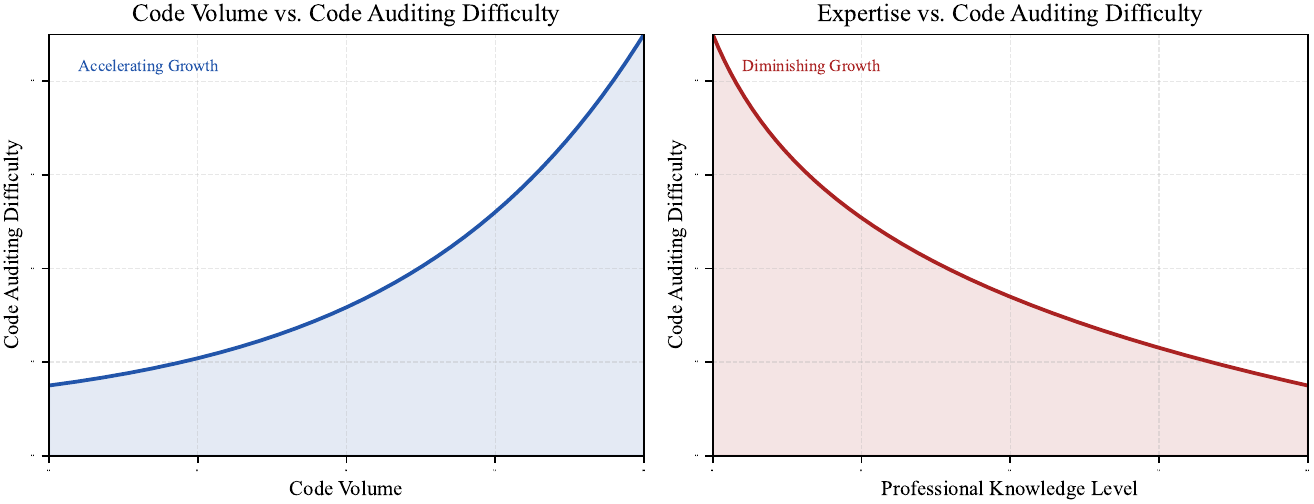}
    \caption{Code auditing difficulty. A larger code volume, coupled with a lower level of professional knowledge, amplifies the complexity of code auditing and consequently increases the code's exposure to poisoning.}
    \label{fig:Code Auditing Difficulty}
\end{figure}

Code hosting platforms are widely used for model development. Three distinct features characterize the rise of ML code hosting platforms. \textbf{Accelerating Development and Lowering Barriers:} The proliferation of open-source code and public repositories (e.g., GitHub~\cite{github}, Hugging Face~\cite{huggingface}) has significantly accelerated ML model development and democratized access. These platforms offer a wealth of pre-trained models, mature frameworks (PyTorch, TensorFlow), algorithm implementations, and toolchains. This ecosystem enables even non-ML experts to build high-performance models on sensitive data, such as medical records~\cite{liu2022loneneuron}. \textbf{Demand for Building Complex Models:} The inherent complexity of modern ML models, particularly ML architectures (encompassing custom loss functions, intricate model structures~\cite{xieintriguing,zada2022pure}, and specialized optimizers), makes building and debugging them from scratch exceptionally challenging. Code hosting platforms lower this barrier through modularity and community contributions, but they also make it harder for users to fully comprehend the entirety and nuances of the code they adopt. \textbf{Plug-and-Play:} ML developers, especially beginners and non-experts, exhibit a strong tendency towards a "plug-and-play" approach with third-party code, often without thorough scrutiny~\cite{mink2023security}. Due to the inherent opacity of complex ML codebases, many specialized components (e.g., custom layers or loss functions) can be functionally "black-box" to non-experts~\cite{wolf2019huggingface}. This opacity creates a fertile ground for stealthy injection of malicious poisoned code that is difficult to detect during routine usage.

Coding agents have significantly enhanced the efficiency of developers (data holders) in model development. Major technology companies have successively rolled out their proprietary agent solutions, including OpenClaw~\cite{Openclaw}, OpenAI's Codex~\cite{Codex}, and Anthropic's Claude Code~\cite{Claude}. The adoption of coding agents for code modification in ML model training has become widespread across both industrial and academic settings. Although these agents are designed to require user authorization prior to executing operations, in practice, users commonly place unconditional trust in coding agents, granting blanket approval without conducting any meaningful review~\cite{qu2026overeagercodingagentsmeasuring}. Recently, coding agents have been reported to introduce code backdoors—whether actively~\cite{ClaudePPPPPP} or passively~\cite{PPPPPPPPPPPP}—which has sparked considerable concern. Effective July 10, 2026, Alibaba imposed a comprehensive ban on all models within Anthropic's Claude family, including the Claude Code programming agent. Likewise, on June 30, 2026, Meta imposed strict restrictions on its engineers' use of Claude Code and Codex, aiming to mitigate the risks associated with model distillation.

Building upon the above context, we construct a code-level property inference attack, which represents a highly innovative and realistic threat model.

\subsection{Machine Learning Models}
Machine Learning (ML) models are trained using labeled data in supervised learning~\cite{caruana2006empirical}. Formally, a training set $ D $ consists of feature vectors $ \mathbf{x} \subseteq {R}^d $ and corresponding labels $ \mathbf{y} \subseteq {R}^m $. The model learns a mapping $ M: \mathbf{x} \to {R}^m $ by employing an optimization algorithm (e.g., Stochastic Gradient Descent~\cite{bottou2012stochastic}) to minimize a predefined loss function. Neural network classifiers, in addition to predicting a class label, typically output a probability distribution over all candidate classes. For a multi-class classification task, given an input $ \mathbf{x} $, the model produces an output vector $ (y_1, \dots, y_m) $ such that $ \sum_{i=1}^{m} y_i = 1 $, where each $ y_i $ denotes the model's confidence (or predicted probability) for the $ i $-th class. This output is computed through successive layers of linear transformations followed by non-linear activation functions. The final predicted class is determined by selecting the class with the highest confidence score. The development of ML models now routinely depends on code hosting platforms and coding agents. We provide a qualitative analysis of code auditing difficulty, as shown in Figure~\ref{fig:Code Auditing Difficulty}. The substantial challenges associated with code auditing, in our view, constitute the foundational premise that validates the real-world feasibility of our code-level PIA.

\subsection{Property Inference Attacks}
\label{PIAs}

\begin{table}[t]
\centering
\caption{Comparison of different methods. CPPIA eliminates the requirements to control the training set or the upstream model (overly permissive assumptions), to train extensive shadow models (prohibitively resource-intensive), or to access model logits. Instead, CPPIA conducts the attack in the most extreme label-only setting and can resist defenses.}
\label{tab:method_comparison}
\adjustbox{width=0.48\textwidth}{
\begin{tabular}{lccccccc}
\toprule
\textbf{Method} & \textbf{Data Access} & \textbf{Model Access} & \textbf{Shadow model} & \textbf{Logits} & \textbf{Defense} & \textbf{Code Access} & \textbf{Label-only} \\
\midrule
Mahloujifar (S\&P~'22)~\cite{mahloujifar2022property} & $\surd$ & $\times$ & $\surd$ & $\surd$ & $\times$ & $\times$ & $\times$ \\
SNAP (S\&P~'23)~\cite{snap}        & $\surd$ & $\times$ & $\surd$ & $\surd$ & $\times$ & $\times$ & $\times$ \\
Tian (CVPR~'23)~\cite{tian2023manipulating}        & $\times$  & $\surd$ & $\surd$ & $\surd$ & $\times$ & $\times$ & $\times$ \\
CPPIA (ours)          & $\times$  & $\times$ & $\times$ & $\times$ & $\surd$ & $\surd$ & $\surd$ \\
\bottomrule
\end{tabular}
}
\end{table}

Property Inference Attacks (PIAs), which are designed to infer global property information of the training set, have witnessed a series of notable advancements. Ateniese \textit{et al.}~\cite{ateniese2015hacking} first systematically formulate the PIA problem against Support Vector Machines (SVMs)~\cite{hearst1998support} and Hidden Markov Models (HMMs)~\cite{eddy1996hidden}. Their attack operates in a white-box setting, with its core mechanism being a meta-classifier trained on multiple shadow models. Subsequently, Ganju \textit{et al.}~\cite{ganju2018property} extend PIAs to the domain of neural networks, focusing specifically on their implementation against fully connected neural networks (FCNNs). To better align with practical attack scenarios, Zhang \textit{et al.}~\cite{zhao2020idlg} propose a PIA scheme under a black-box setting, which still relies on shadow models to construct a meta-classifier. Mahloujifar \textit{et al.}~\cite{mahloujifar2022property} demonstrate that introducing data poisoning attacks~\cite{biggio2012poisoning} into the shadow model training process can significantly enhance the effectiveness of PIAs. In terms of theoretical formalization, Suri and Evans~\cite{suri2022formalizing} are the first to adapt the MIAs cryptographic game framework to provide a rigorous formal definition for PIAs. They further extend the attack on FCNNs to Convolutional Neural Networks (CNNs) in a white-box setting. Zhou \textit{et al.}~\cite{zhou2022property} pioneer a black-box PIA specifically targeting generative models. Furthermore, Chaudhari \textit{et al.}~\cite{snap} propose an efficient data poisoning strategy named SNAP, which relies on only four shadow models and a poisoning ratio $p$ of poisoned samples to conduct property inference. Tian \textit{et al.}~\cite{tian2023manipulating}, during upstream training, introduce an auxiliary loss that causes certain feature dimensions (referred to as "secreting activations") to exhibit controllable magnitude differences between samples with a specific property and those without it. As a result, during downstream fine-tuning, the corresponding parameters are updated on the data containing the property. The adversary can then detect these update differences using white-box or black-box methods to infer properties. Different from data poisoning (data-level)~\cite{mahloujifar2022property,snap} and model poisoning (model-level)~\cite{tian2023manipulating}, our attack, CPPIA, represents the first code-level PIA, as shown in Table~\ref{tab:method_comparison}. In the context where the use of code hosting platforms and coding agents for development has become the norm, CPPIA introduces a new vector for PIAs, as opposed to manipulating either models or data.

\noindent \textbf{Defenses.} To our knowledge, Inf2Guard (USENIX~'24)~\cite{noorbakhsh2024inf2guard} is currently the dedicated defense mechanism proposed to mitigate PIAs. This method draws inspiration from existing neural mutual information estimation techniques, converting the intractable computation of exact mutual information into a tractable variational lower bound. Consequently, model training and inference can rely solely on privacy-processed representations without directly exposing or handling the original sensitive data. Traditional Differential Privacy (DP) technique (DPSGD)~\cite{abadi2016deep} is ineffective against PIAs, as its protection objective is to obfuscate individual sample membership, whereas PIAs aim to infer global statistical properties of the training set~\cite{suri2022formalizing}. Chen \textit{et al.}~\cite{chen2023protecting} propose a framework of mechanisms to defend against PIAs on specific global dataset properties. Their core methodology perturbs query outputs with fine-grained noise, aiming to strike an optimal balance between preserving privacy and maximizing data accuracy across varying data model assumptions. A key instantiation within this framework is the ExpM algorithm, which provides a provably effective and computationally efficient solution for global property privacy. It operates by injecting Laplace or Gaussian noise, where the noise magnitude is precisely calibrated to the expected sensitivity (distance) of the queries, all within well-defined data model constraints.

\section{Threat Model}

\label{sec:Threat}
Developing ML models is a highly specialized task that demands substantial expertise and engineering skills from developers. A mature code framework often comprises tens of thousands of lines of code, making development from scratch rarely the first choice for ML model development. Instead, it has become common practice to leverage abundant, well-written code repositories~\cite{github,huggingface} and coding agents~\cite{Codex,Claude} from third parties for model development~\cite{gu2026spacezonedemystifyingsecurity}. The primary reason is straightforward: utilizing existing codebases and coding agents can significantly shorten the development cycle. Data holders with limited ML expertise can thus develop functional ML models. Moreover, a mature codebase is typically maintained and frequently updated by dozens of contributors, which undoubtedly increases the difficulty of security audits. Adversaries can even impersonate legitimate developers and contribute malicious code. Indeed, real-world incidents of code poisoning attacks have been reported~\cite{wwww,eeee}, spurring considerable research interest. We are the first to investigate how untrusted training code can be exploited to perform PIAs.

\subsection{Adversaries' Goals}
\label{world}

In this section, we outline the goals of PIAs, covering three variants: property inference, property existence attack, and property size estimation.

\noindent \textbf{Property Inference.} We adhere to the definition of PIAs established by SNAP~\cite{snap} and Ateniese \textit{et al.}~\cite{ateniese2015hacking}. A data holder trains a model on a dataset $D$, resulting in a final model $ M: \mathbf{x} \to {R}^m $. The adversary's goal is to infer the proportion of a target property within the model's training set. The attack is formalized as a distinguishing game~\cite{suri2022formalizing}, where the adversary must discriminate between two worlds:
\begin{itemize}
    \item \textbf{World 0}: Model $M$ is trained on samples where the proportion of data possessing the target property $F$ is $t_0$.
    \item \textbf{World 1}: Model $M$ is trained on samples where the proportion of data possessing the target property $F$ is $t_1$.
\end{itemize}
Unlike SNAP~\cite{snap} and Mahloujifar \textit{et al.}~\cite{mahloujifar2022property}, where adversaries submit carefully crafted poisoned samples to the data holder, our privacy game is defined as follows:
\begin{enumerate}
    \item The adversary publicly releases a code repository containing malicious code.
    \item The data holder uses this repository to develop an ML model. They assess the model's performance to determine the code's usability without suspecting potential malicious behavior.
    \item The data holder releases the trained model as a black-box service, providing only predicted labels.
    \item The adversary queries the model with a set of special secret samples. The model's output directly encodes the property information. The adversary decodes this output to infer the property information and win the game.
\end{enumerate}

\noindent \textbf{Property Existence Attack.} This is a special case of property inference where $t_0 = 0$ and $t_1 \ne 0$. The adversary's goal is to determine whether any samples possessing the target property exist in the training set.

\noindent \textbf{Property Size Estimation.} As a generalization of property inference, property size estimation requires the adversary to infer the precise proportion $t$ of the target property, rather than distinguishing between two pre-defined proportions ($t_0$ and $t_1$). SNAP~\cite{snap} often relies on iterative, brute-force searches to approximate the true value of $t$. This incurs substantial computational cost, as the true proportion is a continuous value, and coarse-grained searches may compromise attack performance. In contrast, our proposed attack can directly encode the property information from the training set, theoretically enabling continuous and reliable estimation of the true proportion $t$.

\subsection{Adversaries' Capabilities}
We assume that a malicious code provider is capable of either embedding poisoned code within the published code repository or deploying untrusted coding agents. For non-specialists, inspecting complex, opaque, and sparsely commented code is extremely challenging. Modifications to the code structure may often be justified for legitimate purposes, such as adding auxiliary branches to address dataset imbalance or to enhance adversarial robustness. This very complexity and irregularity make the architecture harder to comprehend and easier for an adversary to exploit.

In contrast to prior works~\cite{mahloujifar2022property,snap}, we do not require the adversary to possess a shadow dataset (from the same distribution) or to incur substantial computational costs for training shadow models. Unlike in SNAP~\cite{snap}, the adversary in our setting does not require control over the data holders' training set; nor does the adversary need full control over the upstream training of the model, as assumed by Tian \textit{et al.}~\cite{tian2023manipulating}. For model access, we adopt the strongest (most restrictive) assumption: the adversary is limited to label-only queries to the model.

\subsection{Data Holders' Capabilities}
Our attack primarily targets data holders who may not have extensive ML expertise. These data holders typically download code from third-party providers or coding agents and use it to train models on their own datasets. To clarify our threat model, we assume that the data holder executes the obtained code within a secure and trusted isolated environment. The data holder does not disclose any information about the training process to the adversary. The adversary is not permitted to directly manipulate the training set. After training, the data holders deploy the model on a platform that only provides black-box query access. We consider a highly restrictive black-box setting where the model returns only the predicted label (top-1 class) for each query, leaking no other information such as confidence scores, logits, or intermediate outputs.

We assume that data holders of ML codebases lack specialized background knowledge or awareness of potential ML attacks, rendering them unable to meticulously review code or identify potential malicious functionalities. This assumption aligns with the common premise in existing code poisoning research~\cite{bagdasaryan2021blind,liu2022loneneuron,song2017machine} and is corroborated by several related user studies in the field~\cite{boenisch2021never,kumar2020adversarial}. Chen \textit{et al.} (NDSS '25)~\cite{chen} demonstrate that membership disclosure can be facilitated in deep learning models by poisoning the training code. Mink \textit{et al.} (USENIX '23)~\cite{mink2023security} point out that practitioners generally have limited awareness of ML attacks and often do not take targeted preventive measures, primarily due to a lack of clear guidelines regarding adversarial ML and related areas. Similar findings have been reported in other studies~\cite{kumar2020adversarial}. In the user survey conducted by Liu \textit{et al.} (CCS '22)~\cite{liu2022loneneuron}, a significant proportion (averaging over 64\%) of participants admitted they do not manually inspect external code before use. Collectively, these findings indicate that typical ML data holders face significant challenges in performing effective code review. Although dedicated code analysis tools for detecting machine learning PIAs are currently lacking, we assume data holders can still actively test the performance of external codebases in two ways: (1) evaluating its accuracy on their domain-specific task, and (2) applying existing privacy defense algorithms (such as those proposed in prior works~\cite{noorbakhsh2024inf2guard,abadi2016deep,chen2023protecting}) to mitigate privacy leakage risks. The former is a common validation practice, while the latter serves as a method for assessing whether an untrusted codebase causes significant privacy harm.

\section{Methodology}
\label{Methodology}
We first outline our design goals in Section \ref{DesignGoals}, followed by a discussion of the challenges associated with achieving these goals in Section \ref{DesignChallenges}. Next, we introduce the underlying attack overview in Section \ref{AttackPrinciple}. Finally, we detail the algorithmic design of CPPIA in Section \ref{AttackApproach}.

\subsection{Design Goals}
\label{DesignGoals}
\noindent \textbf{Goal 1. Achieve 100\% accuracy in property inference attacks.} We aim to perform PIAs with zero error. In privacy-sensitive domains such as healthcare and finance, any inference error is unacceptable. A single incorrect inference can lead to significant decision-making failures, thereby harming both corporate and personal interests. Therefore, a perfect PIA is crucial for privacy regulation and risk management.

\noindent \textbf{Goal 2. Ensure no degradation in model accuracy.} The model compromised via code poisoning must maintain its original high performance. We note that existing PIAs~\cite{mahloujifar2022property,snap,tian2023manipulating} that utilize data-poisoning and model-poisoning to execute property inference often impair model performance. This can easily raise user suspicion, as a poisoned model with lower accuracy is likely to be discarded. Consequently, CPPIA should not affect model performance.

\noindent \textbf{Goal 3. Require minimal computational overhead to execute the attack.} Prior works~\cite{zhang2021leakage,mahloujifar2022property,snap} rely on training shadow models, an approach that demands substantial computational resources and is difficult to deploy in practical scenarios. Although SNAP~\cite{snap} reduces the number of shadow models, it still requires training four shadow models to carry out the attack. Therefore, designing a method with minimal computational cost holds significant practical value.

\noindent \textbf{Goal 4. Evade existing property inference defenses.} We consider it necessary to evaluate attack performance under defensive settings. Our designed attack aims to bypass existing defensive algorithms while still maintaining 100\% attack accuracy under label-only black-box settings.

In our setup, while the adversary can leak property information by tampering with the training code, the data holder can also employ measures to reduce the potential for property privacy leakage. To achieve this, the data holder can decide whether to discard a model based on its accuracy performance. Furthermore, the data holder can leverage existing property inference defenses to mitigate privacy leakage. Neither of these outcomes is desirable for the adversary.

\subsection{Design Challenges}
\label{DesignChallenges}
Although PIAs have been extensively studied, existing works largely fail to achieve the aforementioned four design goals. We analyze their limitations below, which also outline the design challenges our work aims to address.

PIAs are executed by analyzing a model's outputs. Typically, the adversary first trains an ensemble of shadow models to compute a classification threshold. This threshold is then used to extract the property information from the target model's predictions. Recent works~\cite{mahloujifar2022property,snap} focus on leveraging data poisoning to manipulate the model into leaking more private information. However, these prior attacks face four critical limitations.

\begin{enumerate}
    \item Reliance on Imperfect Shadow Models: Threshold-based inference methods are fundamentally constrained by the quality of the shadow models, which cannot guarantee perfect attack accuracy, thereby failing to meet Goal 1 (100\% attack accuracy).
    \item Inevitable Model Accuracy Degradation: Data/model poisoning attacks that directly manipulate model parameters invariably degrade the model's performance on its primary task, raising suspicions from users and violating Goal 2 (degradation of model accuracy).
    \item Prohibitive Architectural and Computational Cost: The shadow model technique requires knowledge of a similar model architecture and a large ensemble of models, incurring massive computational overhead. This contradicts Goal 3 (practical efficiency).
    \item Lack of Evaluation Against Defenses: Due to the nascent stage of property inference defenses, most existing PIAs have not been rigorously tested against them, which is an oversight that ignores Goal 4 (robustness to defenses).
\end{enumerate}

Consequently, the prevailing paradigm of using data poisoning~\cite{mahloujifar2022property,snap} and model poisoning~\cite{tian2023manipulating} to amplify property leakage proves untenable as it simultaneously suboptimizes attack accuracy, model accuracy, efficiency, and defensive robustness. Addressing this holistic set of challenges constitutes the primary contribution of our work.

\begin{figure}[t]
    \centering
    \includegraphics[width=0.48\textwidth]{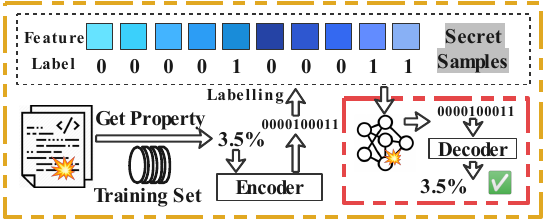}
    \caption{Attack process. The label generation process for secret samples is designed to embed the property information into the model's predictions. Specifically, the poisoned code encodes the target property into a predetermined label sequence, which is then assigned to these samples. After model deployment, the adversary queries the model with the secret samples to obtain this sequence and subsequently decodes it to reconstruct the original property information.}
    \label{fig:ss}
\end{figure}



\begin{figure}[t]
    \centering
    \begin{subfigure}[b]{0.41\textwidth}
        \centering
        \includegraphics[width=\textwidth]{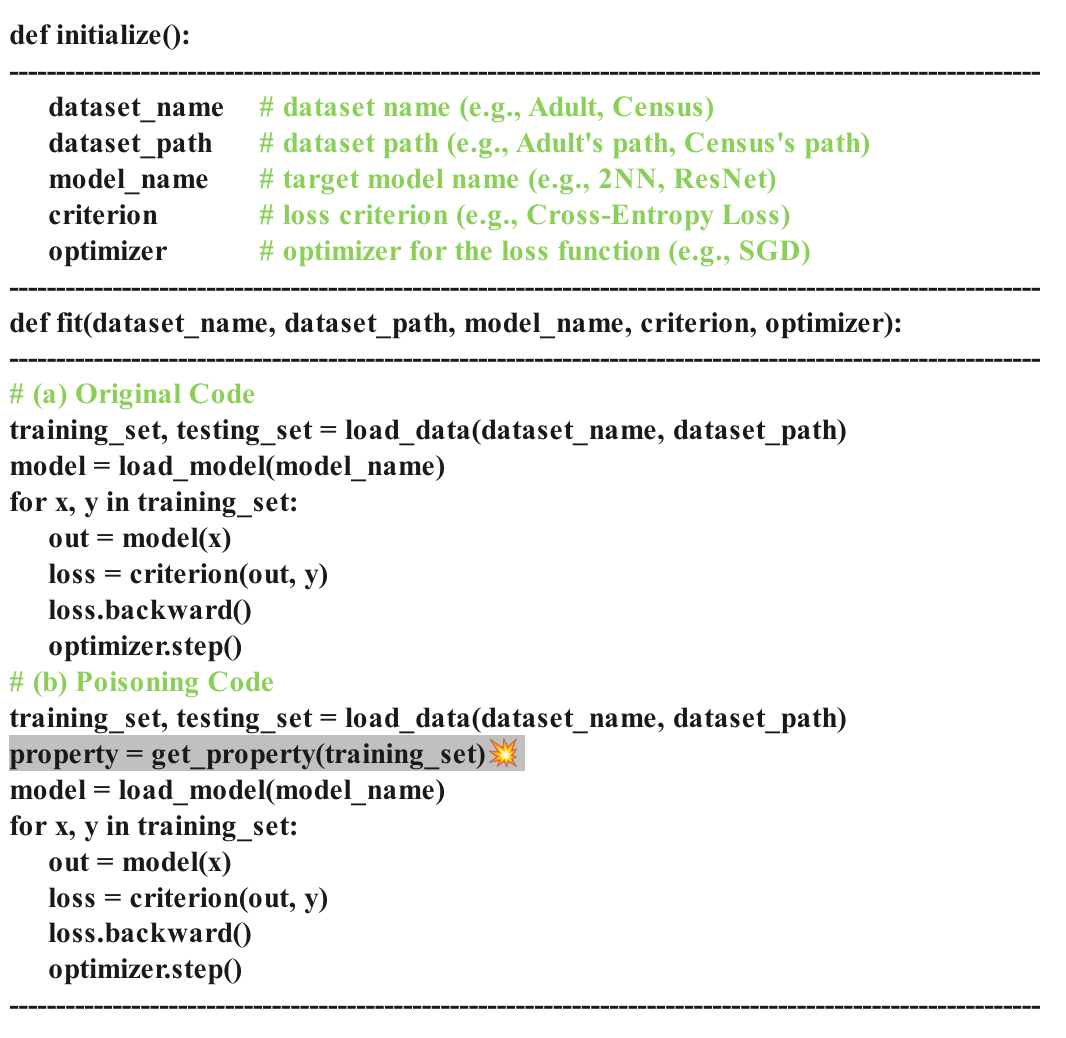}
    \end{subfigure}
    \begin{subfigure}[b]{0.41\textwidth}  
        \centering
        \includegraphics[width=\textwidth]{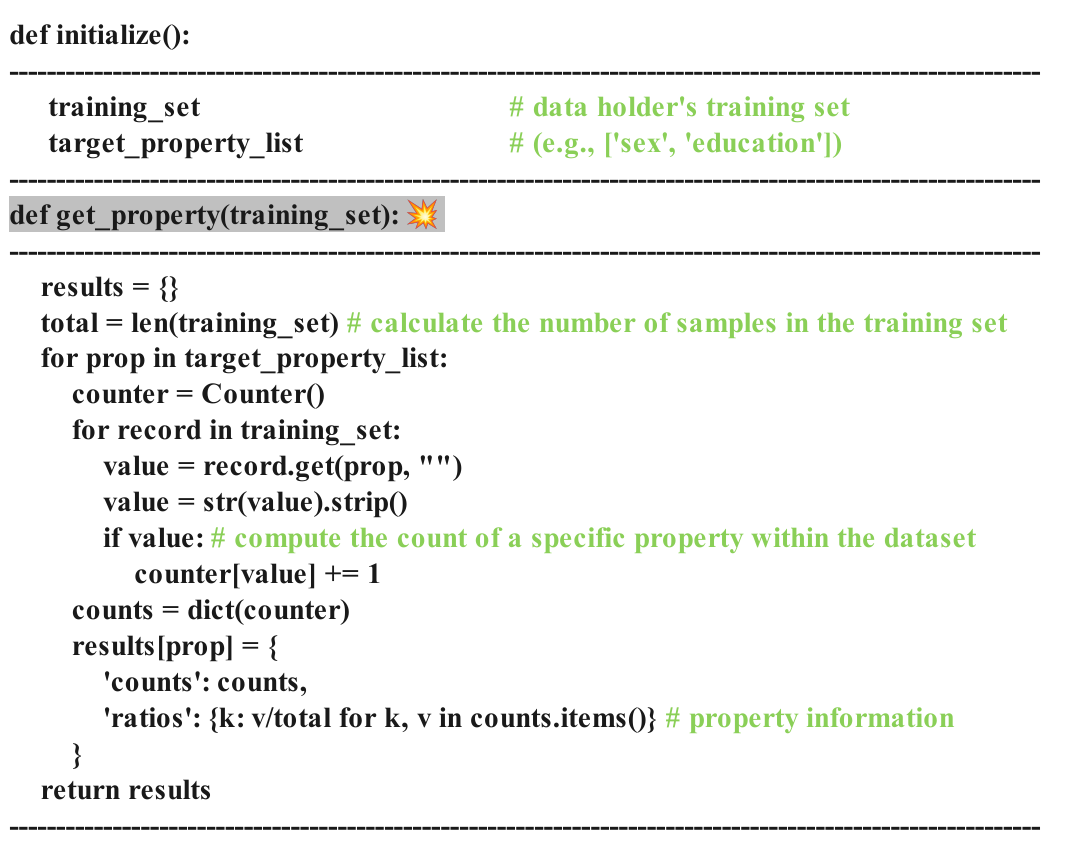}
    \end{subfigure}
    \caption{The code for computing statistical property information.}
    \label{fig:code1}
\end{figure}

\begin{figure}[t]
    \centering
    \includegraphics[width=0.41\textwidth]{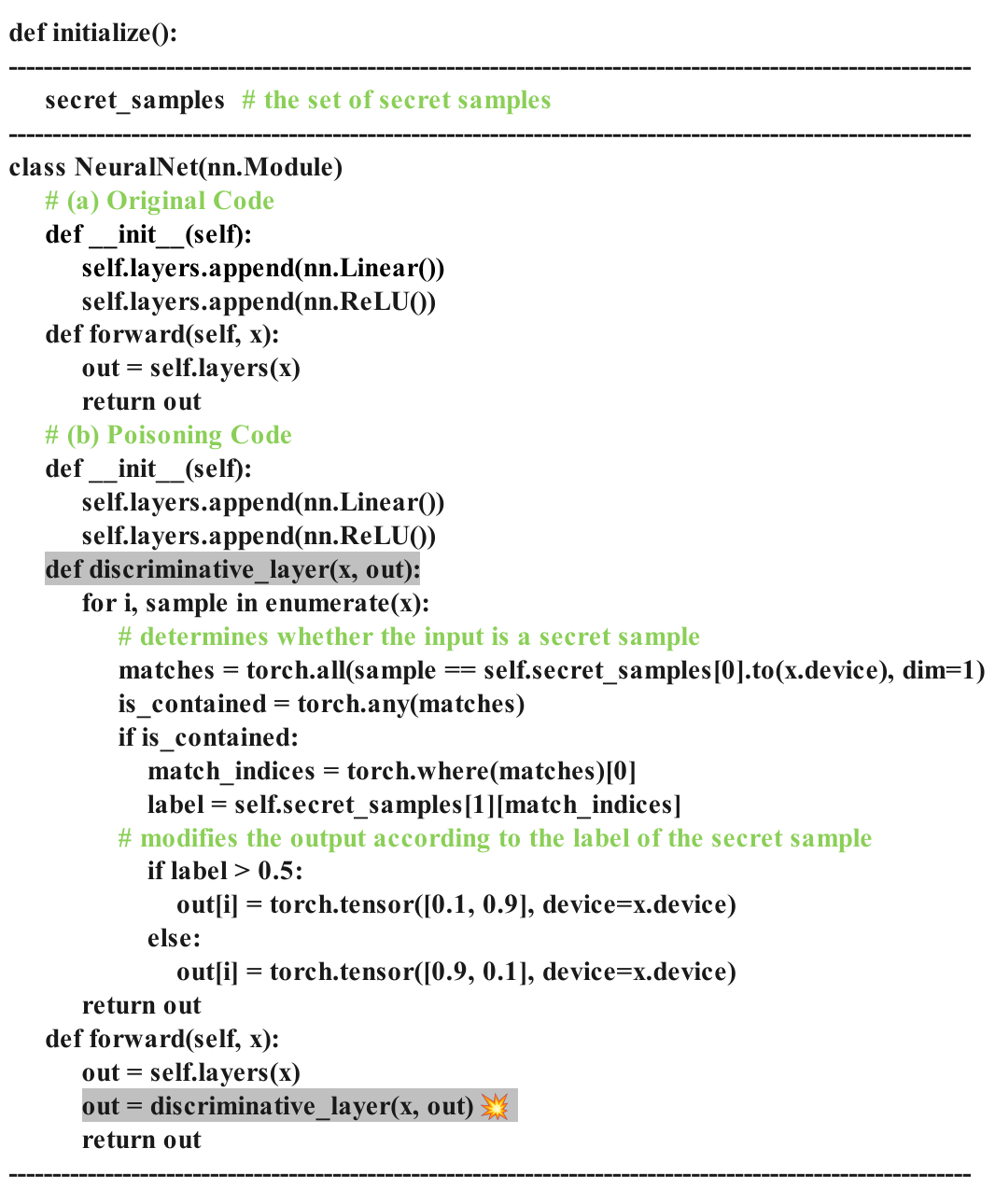}
    \caption{A $discriminative\_layer$() is embedded within the poisoned code, with the explicit purpose of achieving high-accuracy predictions on the secret samples. Since it is not part of the model architecture, it cannot be discovered by using $print(model)$.}
    \label{fig:code3}
\end{figure}

\subsection{Attack Overview}
\label{AttackPrinciple}
The underlying overview of CPPIA can be described as the decoupling and transfer of property information, as shown in Figures~\ref{fig:Overview},~\ref{fig:ss}. For the dataset provided by the data holder, the adversary can use poisoned code to first statistically compute the target property information from this dataset. This privacy-sensitive information is then covertly transferred and embedded into the model outputs corresponding to a set of specially crafted secret samples (outlier samples, as shown in Table \ref{tab:data_compact}). Once the model is published as an API, the adversary can query the model using these known secret samples to retrieve the embedded property information. This process incurs negligible computational cost for the adversary and, in theory, enables a perfect attack with 100\% accuracy.

\begin{algorithm}[t]
\caption{CPPIA Algorithm}
\begin{algorithmic}[1]
\label{alg_CPPIA}
\STATE \textbf{Symbol:} secret samples $(x_{sec_1}$,$y_{sec_1})$,$\dots$,$(x_{sec_{10}}$,$y_{sec_{10}})$, dataset $D$, target property $t$, code-poisoned model $\hat{M}$
\STATE \textbf{Upload:} Adversaries upload the poisoned code.
\STATE \textbf{Download:} Data holders download the poisoned code and train models in the trusted
isolated environment.
\STATE \COMMENT{Poisoned code is conducted during \textbf{normal training}.}
\STATE \COMMENT{\textbf{Extract property information.}}
\STATE $t\gets get\_property(D)$
\STATE \COMMENT{\textbf{Encode the target property into secret samples.}}
\STATE $ y_{sec_{1}}, \dots, y_{sec_{10}} \gets Encoder(t)$
\STATE \COMMENT{\textbf{Embed the secret samples into the model.}}
\STATE \COMMENT{Release the API (label-only) access of $\hat{M}$.}
\STATE $ API(\cdot) \gets \hat{M} \gets (x_{sec_1},y_{sec_1}),\dots,(x_{sec_{10}},y_{sec_{10}})$
\STATE {\textbf{Inference:}} Adversaries conduct attacks to obtain $t$.
\STATE $ (y^*_{sec_{1}}, \dots, y^*_{sec_{10}}) \gets API(x_{sec_{1}}, \dots, x_{sec_{10}})$
\STATE $ t \gets Decoder(y^*_{sec_{1}}, \dots, y^*_{sec_{10}})$
\end{algorithmic}
\end{algorithm}

\subsection{Attack Details}
\label{AttackApproach}
Following the training code's loading of the dataset, we first perform a statistical profiling of its properties, as illustrated in Figure \ref{fig:code1}. Compared to the original code, the poisoned code incorporates a step to statistically analyze the dataset immediately after loading it. This enables the adversary to decouple and preprocess the private properties of the dataset for subsequent attack stages. For clarity in exposition, we have explicitly named this function $get\_property$(). This function takes the training set and the target property as inputs and returns the statistical property. In a practical attack, it can be disguised under benign names such as $data\_augmentation$() to obfuscate its true intent. The goal of PIAs is precisely to leak this statistical information. Existing PIAs~\cite{mahloujifar2022property,snap,zhou2022property} rely on training shadow models to compute classification thresholds for estimating the model's properties. These methods not only demand substantial computational resources but also yield suboptimal attack accuracy. Conversely, property inference defenses aim to minimize the leakage of this information, making different property information indistinguishable. In this work, we consider an extremely strict black-box setting, where the adversary can only access the model's final output labels (label-only).

To induce the model to leak private information, we encode the property information into a binary sequence and assign it as labels to a set of secret samples (kept hidden from the data holder). The adversary's goal is to make the model precisely output this corresponding binary sequence when queried with those secret samples. This sequence is subsequently decoded to recover the original property information of the model, as illustrated in Algorithm~\ref{alg_CPPIA}. Through this mechanism, the data holder's property information is effectively embedded into the model's outputs for a set of secret samples (known only to the adversary). These samples are generated during training by the adversary's poisoned code and are co-trained alongside the data holder's training set.

The secret samples planted by our poisoned code must satisfy two necessary conditions. First, the training model must be compelled to output their encoded labels with high fidelity. Our key insight is that outlier samples (e.g., a sample with "Husband" despite having an "Age" property of 5) are more susceptible to memorization and less prone to interference during training. Building on this observation, we encode the binary sequence of property information into a set of labels for such outlier samples. These intrinsically contradictory samples are unlikely to appear in a user's normal testing set, yet the adversary can leverage them to infer the property information with high precision. Second, the encoding must have a strict, one-to-one correspondence with the target property and be reliably decodable. In our implementation, we utilize ten outlier samples to encode the property value, achieving a numerical precision of up to three decimal places, as illustrated in Table~\ref{tab:data_compact}.

To illustrate our attack process concretely, we provide a simplified example. Suppose the data holder's training set contains 3.5\% (0.035) of samples with the joint property (Gender=Female, Occupation=Sales). Our attack proceeds as follows:
\begin{enumerate}
    \item Property Information Extraction: The $get\_property$() function first extracts this ratio: 0.035.
    \item Value Encoding: This value is scaled by 1000 to obtain an integer: 35.
    \item Binary Conversion: The integer 35 is converted to its binary representation: $100011_2$.
    \item Padding: The binary string is left-padded with zeros to a fixed length of 10 bits, yielding the final encoded label sequence: $0000100011_2$.
    \item Injection: This sequence $0000100011_2$ is then assigned as the labels for our ten secret outlier samples, which are embedded into the training set by the poisoned code.
\end{enumerate}
Once the model is deployed, the adversary queries it with these ten secret samples. The model's predicted label sequence ($0000100011_2$) is then decoded (by reversing steps 2-4) to recover the original property value of 0.035.

The specific encoding rule described above is not the only viable strategy. The adversary can customize the encoding/decoding scheme (e.g., using different bases, error-correcting codes, or cryptographic primitives) based on the target model and dataset characteristics. Since the choice of encoding does not fundamentally affect the attack accuracy (which depends on the model accuracy in predicting the labels), we adopt this straightforward decimal-to-binary conversion for clarity and simplicity. Our encoding scheme's precision is determined by the capacity of the secret sample set. Since a 10-bit binary sequence can represent values up to 1023, using ten secret samples provides sufficient capacity to encode a decimal value with three decimal places of granularity (by scaling the value by 1000). Correspondingly, fourteen secret samples (14 bits) can support encoding with four decimal places of precision. Furthermore, the encoding rule can be simplified for models with a suitable output space. For instance, if the target model is an n-class classifier (e.g., n=10), we can directly use the predicted class label of a single secret sample to represent a digit from 0 to n-1. This allows us to encode a multi-digit number directly and more efficiently using a sequence of secret samples, each corresponding to one digit. Discussion of more complex, cryptographically involved encoding schemes is beyond the scope of this work.

After the poisoned code embeds the property information into the secret samples, the adversary must ensure that the model correctly outputs their designated labels. Our initial method is to simply inject these secret samples into the training set, expecting the model to learn their labels. However, we quickly identify a critical problem with this naive method. This method presupposes a roughly uniform distribution of classes in the training set, which can make assigning specific labels to the secret samples relatively straightforward. In practice, however, this precondition is highly unrealistic for an adversary. The severe class imbalance in common real-world datasets, such as Adult~\cite{data} (76\% vs. 24\% for classes 0 and 1) and Census~\cite{data} (94\% vs. 6\%), makes it exceedingly difficult to force the model to memorize arbitrary labels for secret samples through standard co-training alone. 

To overcome this fundamental limitation, we introduce a discriminative function from a model output perspective in Figure \ref{fig:code3}. The discriminative function is designed to explicitly guide the predictions for the secret samples. Crucially, this design does not interfere with the processing of benign samples, thus preserving the model's original accuracy. Finally, because the secret samples are known only to the adversary and the stolen property information is embedded solely within the model's outputs for these samples, our attack remains imperceptible to the end user, evading detection. 
In pursuit of enhanced performance, such as improved adversarial robustness~\cite{xieintriguing} or better handling of imbalanced classification~\cite{zada2022pure}, ML researchers have proposed a proliferation of customized architectural modifications (e.g., adding specialized normalization layers). Consequently, an ostensibly "irregular" architectural change may be plausibly property to benign performance-tuning motives. This prevalent practice provides a natural camouflage for our attack model augmentation, allowing it to blend seamlessly into the modern ML development workflow. Refer to Section~\ref{Discussion} for details on the stealthiness of code poisoning and the encoding rules.

\section{Evaluation}
Section~\ref{ExperimentalSetup} elaborates on our experimental setup. Following this, we present the experimental results from Section~\ref{g1} to Section~\ref{de} corresponding to the four goals defined in the threat model. Within Section~\ref{EstimatingSizeofTargetProperty}, we evaluate the performance of property size estimation. Finally, the impact of model complexity is discussed in Section~\ref{modell}. Section~\ref{iomn} examines the impact of the number of models. Section~\ref{iops} investigates the impact of property size. Section~\ref{Additional} describes additional task models.

\begin{table}[t]
\centering
\caption{Target properties considered in Adult, Census, Bank Marketing, and CelebA. The adversary’s goal is to distinguish between the two percentages of the target property shown in the last column.}
\label{attack_type}
\begin{adjustbox}{width=0.45\textwidth}
\begin{tabular}{l|l|l|c}
\toprule
Property Size & Dataset & Target Properties & $t$ \\
\midrule
\multirow{3}{*}{Large} & Adult & Workclass = Private & 20\% vs 40\% \\
& & Race = White; Gender = Male & 15\% vs 30\% \\
\cmidrule(lr){2-4}
& Census & Race = Black & 10\% vs 25\% \\
& & Gender = Female & 30\% vs 50\% \\
\cmidrule(lr){2-4}
\multirow{4}{*}{\textbf{}} & Bank Marketing & Month = May & 10\% vs 25\% \\
& & Marital-Status = Married & 25\% vs 50\% \\
\cmidrule(lr){2-4}
& CelebA & Gender = Male & 30\% vs 70\% \\
& & Age = Old & 25\% vs 60\% \\
& & Wearing Earrings & 15\% vs 40\% \\
\cmidrule(lr){1-4}
\multirow{3}{*}{Medium} & Adult & Gender = Female; Occupation = Sales & 1\% vs 3.5\% \\
& & Marital-Status = Divorced; Gender = Male & 1\% vs 5\% \\
\cmidrule(lr){2-4}
& Census & Education = Bachelors & 2\% vs 8\% \\
& & Industry = Construction & 2\% vs 7\% \\
\cmidrule(lr){2-4}
& Bank Marketing & Contact = Telephone & 1\% vs 6\% \\
& & Previous Campaign = Failure & 3\% vs 8\% \\
\cmidrule(lr){1-4}
\multirow{3}{*}{Small} & Adult & Native Country = Germany & 0\% vs 0.10\% \\
& & Occupation = Protective Services & 0\% vs 0.05\% \\
\cmidrule(lr){2-4}
& Census & Hispanic-Origin = Cuban & 0\% vs 0.20\% \\
\bottomrule
\end{tabular}
\end{adjustbox}
\end{table}

\begin{table}[!t]
\centering
\caption{Ten secret samples for Adult. These samples are generally not employed in real-world applications.}
\label{tab:data_compact}
\begin{adjustbox}{width=0.45\textwidth}
\begin{tabular}{@{\extracolsep{2pt}}ccccccccccccc@{}}
\toprule
age & workclass & edu & marital & occup & rel & race & gender & gain & loss & hrs & country & class \\
\midrule
3 & Private & 20 & Married & Farming & Child & White & F & 435332 & 20589 & 109 & US & 0 \\
104 & ? & 29 & Separated & Handlers & Husband & White & F & 312524 & 28073 & 168 & US & 0 \\
7 & Private & 1 & Married & Service & Child & White & M & 107944 & 31681 & 152 & US & 0 \\
81 & ? & 22 & Divorced & Transport & Husband & Black & M & 256772 & 12855 & 136 & US & 0 \\
1 & Private & 10 & Married & Craft & Child & Black & F & 856134 & 28413 & 103 & US & 0 \\
8 & Private & 23 & Married & ? & Unmarried & Black & M & 681302 & 41889 & 3 & US & 1 \\
99 & Private & 2 & Divorced & Tech & Husband & White & F & 427236 & 15330 & 107 & US & 0 \\
2 & Private & 24 & Never & Service & Child & White & M & 200638 & 29133 & 86 & US & 1 \\
5 & ? & 25 & Never & ? & Wife & White & F & 259142 & 12990 & 164 & US & 0 \\
142 & Local-gov & 30 & Separated & Prof & Wife & Asian & M & 207419 & 20584 & 132 & US & 0 \\
\bottomrule
\end{tabular}%
\end{adjustbox}
\end{table}

\subsection{Experimental Setup}
\label{ExperimentalSetup}

We conduct a comprehensive experimental evaluation of CPPIA using four datasets, eight model architectures, eighteen distinct properties, and three defenses. All experimental results are reported as the mean over five independent runs.

To ensure a direct and fair comparison, we align our experimental setup with SNAP~\cite{snap}, utilizing the same four benchmark datasets: Adult~\cite{data}, Census~\cite{data}, Bank Marketing~\cite{data}, and CelebA~\cite{data2}. Furthermore, we adopt the identical set of target properties to guarantee a like-for-like comparison, as shown in Table~\ref{attack_type}. For the critical poisoning parameter $p$ (the ratio of poisoned samples) in SNAP~\cite{snap}, we strictly adhere to its original configuration as reported in their work. The works of Mahloujifar \textit{et al.}~\cite{mahloujifar2022property} and Tian \textit{et al.}~\cite{tian2023manipulating} are also employed as baselines in our experiments. It is important to clarify that our experimental setup is fair: both Mahloujifar \textit{et al.}~\cite{mahloujifar2022property} and SNAP~\cite{snap} assume that the adversary can inject the generated poisoned samples into the training set, whereas Tian \textit{et al.}~\cite{tian2023manipulating} assume full control over the upstream training process of the target model, which in turn leads to property leakage. By contrast, CPPIA merely involves releasing poisoned code to code hosting platforms and does not assume direct control over the model training process (dataset collection, hyperparameter configuration, internal model parameters, etc.). The datasets employed in our experiments are listed as follows:

\begin{itemize}
    \item \textbf{Adult}. The UCI Adult dataset~\cite{data} is a binary classification dataset containing 48,842 records extracted from the 1994 US Census database. Each record comprises 14 demographic and employment-related properties, such as gender, race, and marital status. The classification task is to predict whether an individual's annual income exceeds \$50,000. The class distribution is imbalanced, with class 0 ($\le $\$50K) accounting for 76\% and class 1 ($>$\$50K) for 24\%. For our primary analysis, we employ a neural network with two hidden layers (32 and 16 neurons, respectively).
    \item \textbf{Census}. The US Census Income dataset~\cite{data} is an extended version of the UCI Adult. It consists of 299,285 records with 41 properties. The classification task is identical to that of Adult. The class distribution is highly imbalanced, with 94\% of samples belonging to class 0 and only 6\% to class 1. We analyze this dataset using the same two-layer neural network architecture as for Adult.
    \item \textbf{Bank Marketing}. The Bank Marketing dataset~\cite{data} is a binary classification dataset comprising 45,211 records. Each record is described by 16 properties. The task is to predict whether a client will subscribe to a term deposit. The data exhibits a significant class imbalance: 88\% for class 0 (no subscription) and 12\% for class 1 (subscription). We utilize the same two-layer neural network architecture as described for Adult.
    \item \textbf{CelebA}. The CelebA dataset~\cite{data2} contains 202,599 celebrity face images, each annotated with 40 binary properties such as gender, race, and whether the person is wearing glasses. The dataset is class-balanced for most properties. In our experiments, we employ ResNet-18~\cite{he2016deep} with the classification target set to predicting whether the person is smiling. This task serves as a fine-grained facial expression recognition problem, leveraging the representational power of ResNet-18~\cite{he2016deep} to capture relevant features from facial regions.
\end{itemize}

For our defense configuration, we implement the three mechanisms evaluated in Section~\ref{PIAs}: Inf2Guard~\cite{noorbakhsh2024inf2guard}, ExpM~\cite{chen2023protecting}, and DPSGD~\cite{abadi2016deep}. The key defense parameter (privacy budget $\epsilon$) for each method is detailed alongside the corresponding experimental results in the following section. A smaller $\epsilon$ provides stronger theoretical privacy guarantees.

\noindent \textbf{Selection of Target Properties.} Table~\ref{attack_type} presents the 18 target properties we have defined across four datasets. Similar to SNAP~\cite{snap}, we categorize the property sizes into three classes: large, medium, and small. Among these, the large and medium properties are used for property inference attacks, while the small properties are employed for property existence attacks.

\noindent \textbf{Secret Samples.} In Table~\ref{tab:data_compact}, we present 10 secret samples constructed from Adult~\cite{data}. These samples cannot exist in the real world, as each sample contains contradictory property values. For instance, a sample with an age of 3 cannot have 20 years of education. As a result, ordinary users typically struggle to query the model using these secret samples. Adversaries covertly embed these samples into the model, allowing adversaries to perform accurate queries. Secret samples are flexibly determinable by the adversary in accordance with the encoding rules.

\noindent \textbf{Metric.} Similar to prior works~\cite{mahloujifar2022property,snap}, the success of the attack is quantified as the attack accuracy (Attack ACC) with which the adversary can correctly identify the proportion of the target property used in the training of the ML model. Model accuracy (Model ACC) reflects whether the model's predictions are correct and is intended for evaluating model performance. Additional task models are discussed in Section~\ref{Additional}.

\subsection{Attack Accuracy (Goal 1)}
\label{g1}

\begin{figure*}[t]
\centering
\begin{subfigure}[b]{0.18\textwidth}
    \centering
    \includegraphics[width=\textwidth]{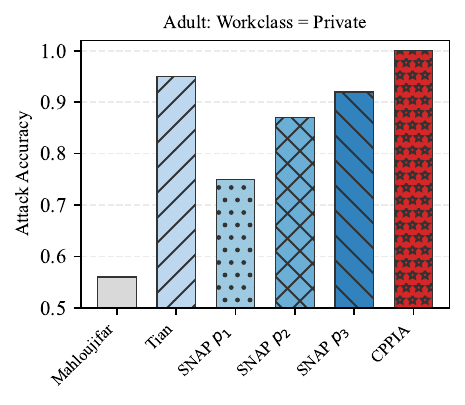}
\end{subfigure}
\hspace{0.01\textwidth}
\begin{subfigure}[b]{0.18\textwidth}
    \centering
    \includegraphics[width=\textwidth]{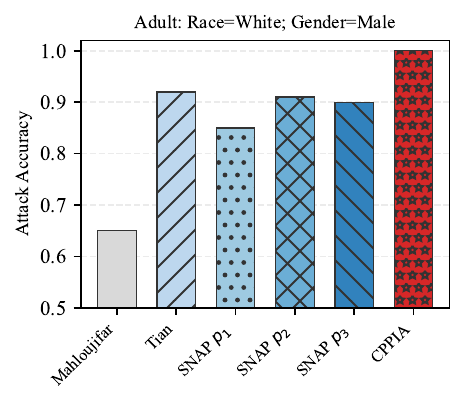}
\end{subfigure}
\hspace{0.01\textwidth}
\begin{subfigure}[b]{0.18\textwidth}
    \centering
    \includegraphics[width=\textwidth]{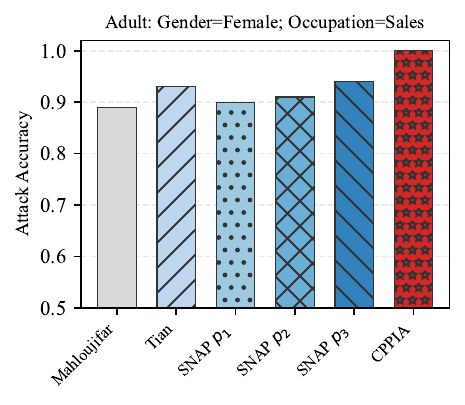}
\end{subfigure}
\hspace{0.01\textwidth}
\begin{subfigure}[b]{0.18\textwidth}
    \centering
    \includegraphics[width=\textwidth]{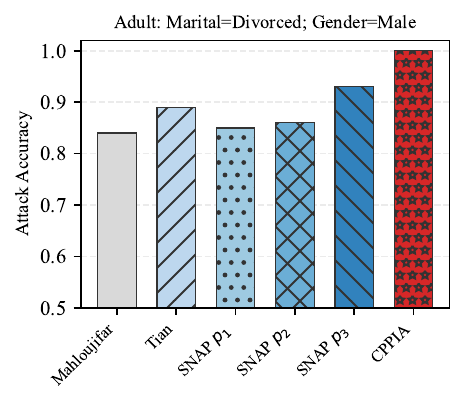}
\end{subfigure}
\hspace{0.01\textwidth}
\begin{subfigure}[b]{0.18\textwidth}
    \centering
    \includegraphics[width=\textwidth]{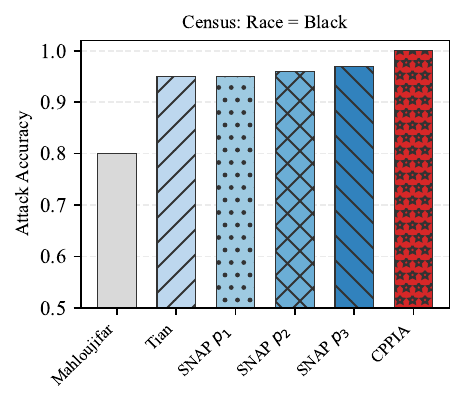}
\end{subfigure}

\vspace{4pt}
\begin{subfigure}[b]{0.18\textwidth}
    \centering
    \includegraphics[width=\textwidth]{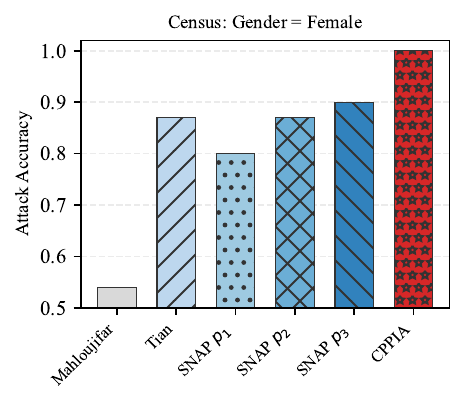}
\end{subfigure}
\hspace{0.01\textwidth}
\begin{subfigure}[b]{0.18\textwidth}
    \centering
    \includegraphics[width=\textwidth]{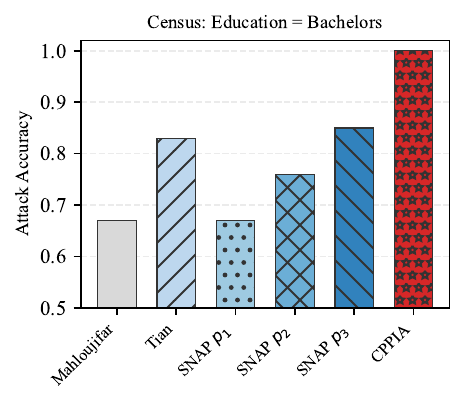}
\end{subfigure}
\hspace{0.01\textwidth}
\begin{subfigure}[b]{0.18\textwidth}
    \centering
    \includegraphics[width=\textwidth]{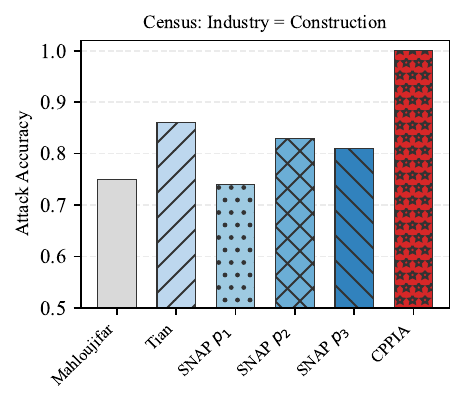}
\end{subfigure}
\hspace{0.01\textwidth}
\begin{subfigure}[b]{0.18\textwidth}
    \centering
    \includegraphics[width=\textwidth]{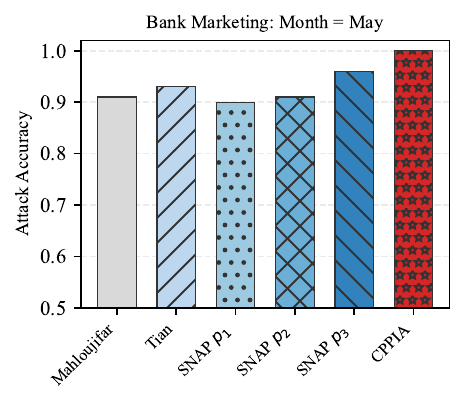}
\end{subfigure}
\hspace{0.01\textwidth}
\begin{subfigure}[b]{0.18\textwidth}
    \centering
    \includegraphics[width=\textwidth]{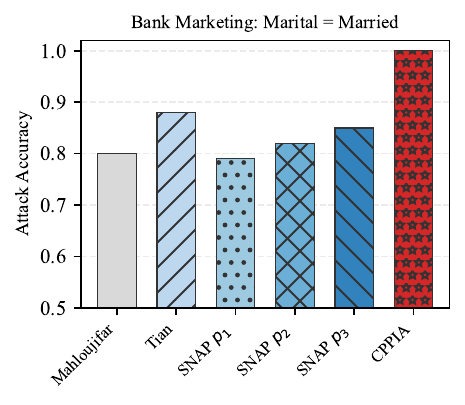}
\end{subfigure}

\vspace{4pt}
\begin{subfigure}[b]{0.18\textwidth}
    \centering
    \includegraphics[width=\textwidth]{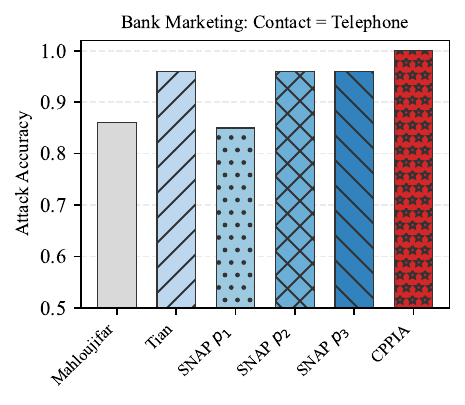}
\end{subfigure}
\hspace{0.01\textwidth}
\begin{subfigure}[b]{0.18\textwidth}
    \centering
    \includegraphics[width=\textwidth]{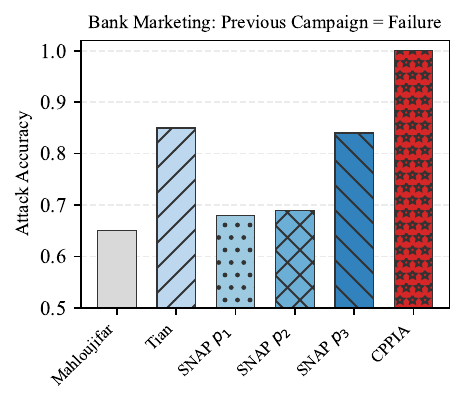}
\end{subfigure}
\hspace{0.01\textwidth}
\begin{subfigure}[b]{0.18\textwidth}
    \centering
    \includegraphics[width=\textwidth]{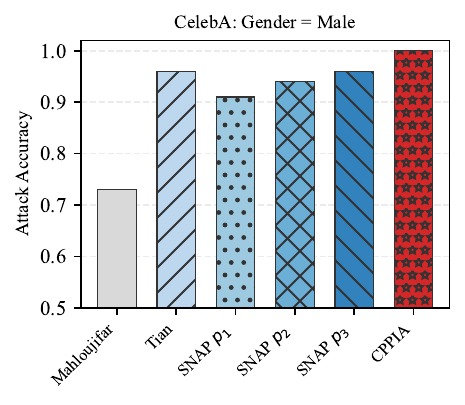}
\end{subfigure}
\hspace{0.01\textwidth}
\begin{subfigure}[b]{0.18\textwidth}
    \centering
    \includegraphics[width=\textwidth]{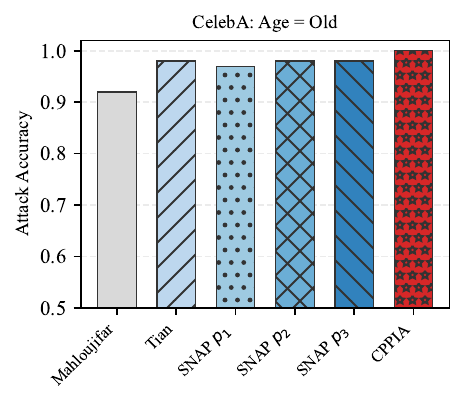}
\end{subfigure}
\hspace{0.01\textwidth}
\begin{subfigure}[b]{0.18\textwidth}
    \centering
    \includegraphics[width=\textwidth]{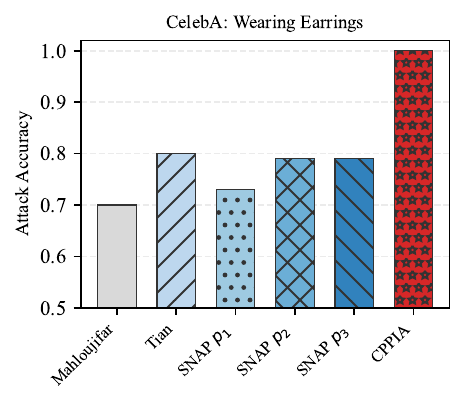}
\end{subfigure}


\caption{Attack accuracy of six inference attacks on each target property across four datasets. Each small subplot corresponds to one target property. Different hatching patterns and colors distinguish attack methods. CPPIA achieves perfect attack accuracy (100\%) on all tasks.}
\label{tab:target_properties}
\end{figure*}

\begin{figure}[t]
    \centering
    \includegraphics[width=0.44\textwidth]{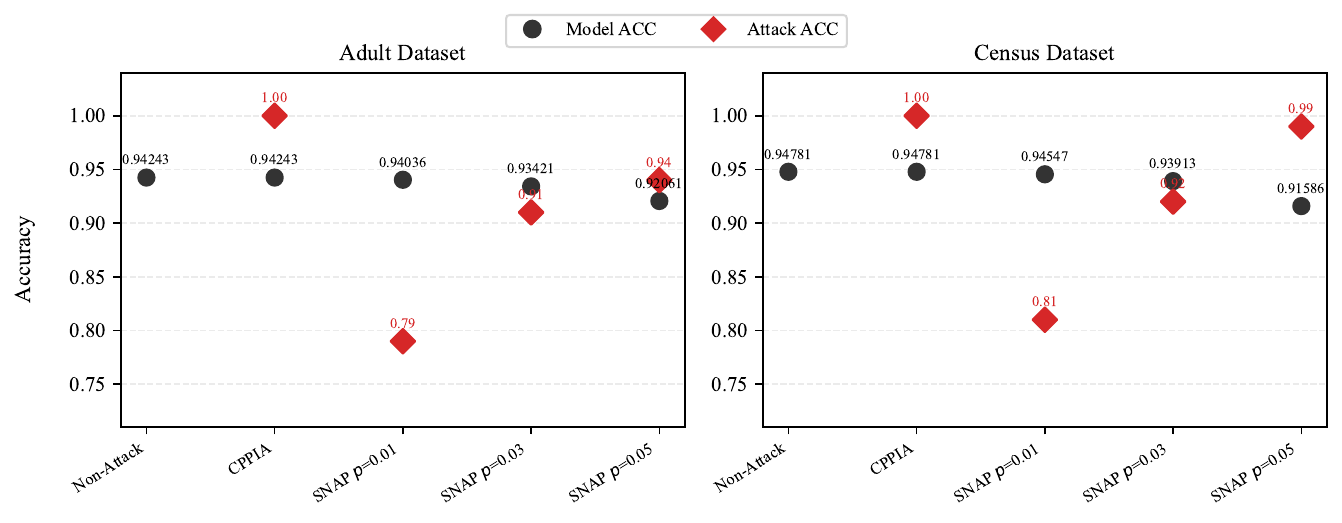}
    \caption{Model accuracy and attack accuracy under four PIAs.}
    \label{fig:Model accuracyModel accuracy}
\end{figure}


\begin{figure}[t]
    \centering
    \includegraphics[width=0.44\textwidth]{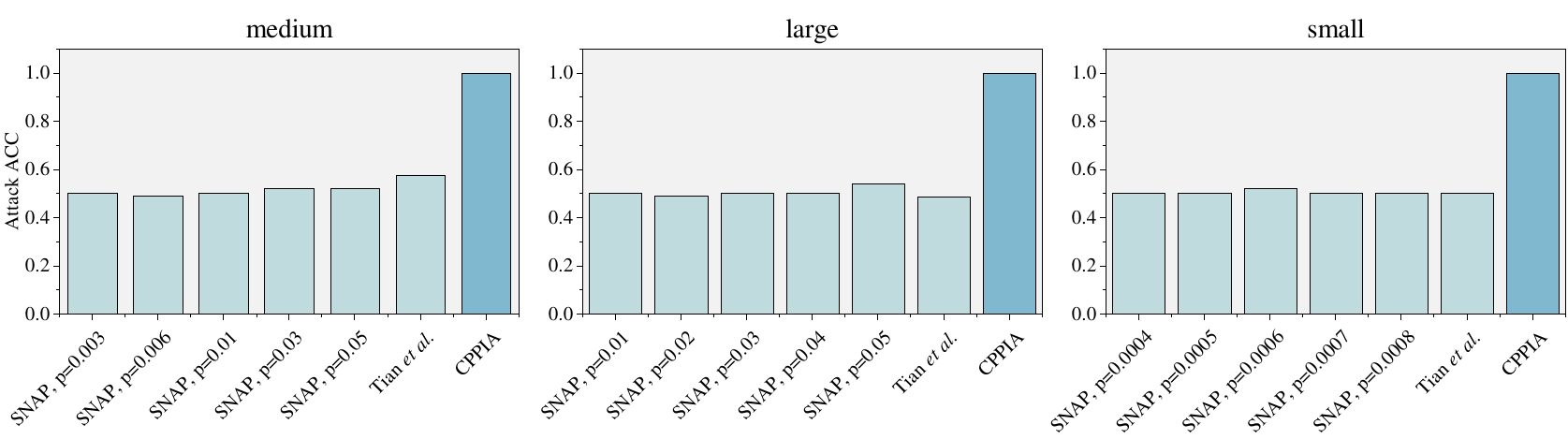}
    \caption{Attack accuracy on Adult under the Inf2Guard defense.}
    \label{fig:7}
\end{figure}

Figure~\ref{tab:target_properties} presents the attack accuracy of CPPIA across eighteen target properties. For SNAP~\cite{snap}, we consider three different settings of the parameter $p$. Our attack consistently achieves a perfect 100\% attack accuracy. By accurately decoding the predicted labels of the secret samples, the adversary can perfectly reconstruct the statistical properties of the data holder's original training set. This complete privacy breach is demonstrated to be robust across a wide range of conditions, including different model architectures (Section~\ref{modell}), datasets, target properties, and property sizes. Notably, CPPIA's 100\% attack accuracy is achieved without any hyperparameter tuning. In contrast, SNAP~\cite{snap} is highly sensitive to critical parameters, such as the poisoning ratio $p$. A detailed comparative analysis with SNAP under varying parameter settings is provided in Section~\ref{de}.

\subsection{Model Accuracy (Goal 2)}
\label{g2}
Figure~\ref{fig:Model accuracyModel accuracy} presents the model accuracy and attack accuracy on Adult and Census. The non-attack model achieves an accuracy of 0.94243. Remarkably, CPPIA attains a perfect 100\% attack accuracy while preserving the model accuracy entirely. Consequently, the data holder cannot unmask CPPIA by monitoring model performance, as the attack induces zero accuracy degradation. In contrast, for SNAP~\cite{snap}, the optimal accuracy of 0.94 is achieved only at a poisoning ratio of $p$=0.05, and it comes at a high cost: the model accuracy drops by 2.182\%. This noticeable performance decline easily alerts the data holder, likely leading to model rejection. For SNAP~\cite{snap}, increasing $p$ to boost attack accuracy inevitably further degrades model accuracy, thereby increasing its detectability. In summary, these results conclusively demonstrate that CPPIA successfully fulfills goal 2: preserving the model accuracy while achieving a perfect attack accuracy.

\subsection{Computational Cost (Goal 3)}
Existing PIAs~\cite{ganju2018property,mahloujifar2022property} require training hundreds of shadow models to mount an attack. SNAP~\cite{snap} requires training four shadow models. To quantify this overhead, our reproduction of SNAP~\cite{snap} across 50 models on a single NVIDIA RTX 4090 GPU revealed that shadow model training alone consumed 4/104 of the total computational budget. As model architectures grow more diverse and complex, this cost becomes prohibitively expensive. Moreover, in practical scenarios, constructing effective shadow models requires the adversary to collect auxiliary datasets that follow a distribution similar to the private data. This endeavor incurs a high additional cost. In stark contrast, CPPIA achieves PIA with virtually no computational overhead. This claim is substantiated by our experiments: the attack simply requires querying the model with a handful of samples and decoding the outputs, an operation that is negligible in cost compared to model training.

\subsection{Robustness under Defenses (Goal 4)}
\label{de}

\begin{figure}[t]
    \centering
    \begin{subfigure}[b]{0.48\textwidth}
        \centering
        \includegraphics[width=\textwidth]{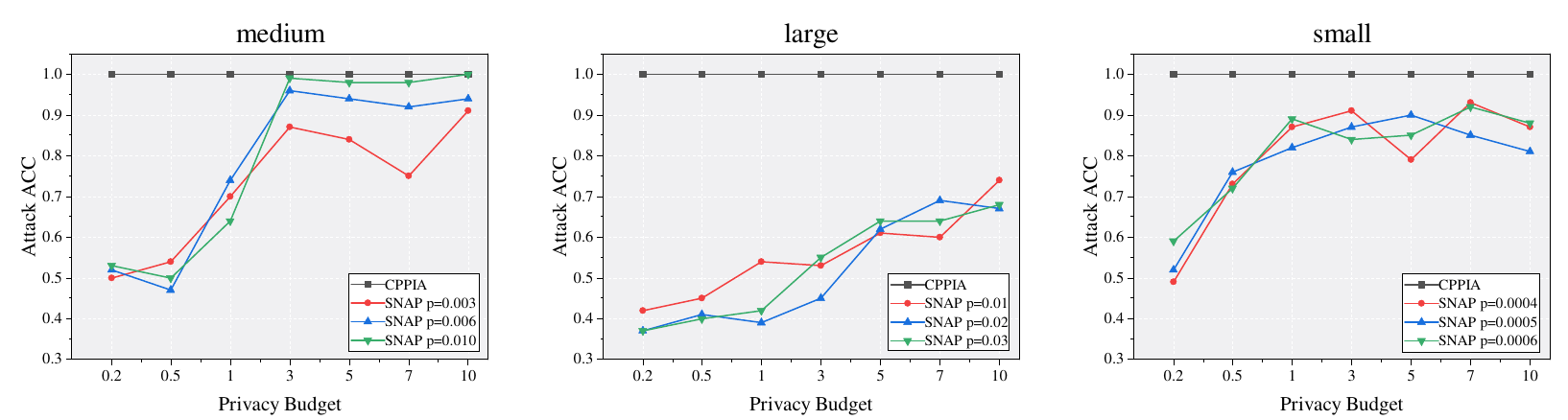}
        \caption{Attack accuracy on Adult under the ExpM defense.}
        \label{fig:8}
    \end{subfigure}
    \hspace{0.01\textwidth}
    \begin{subfigure}[b]{0.48\textwidth}
        \centering
        \includegraphics[width=\textwidth]{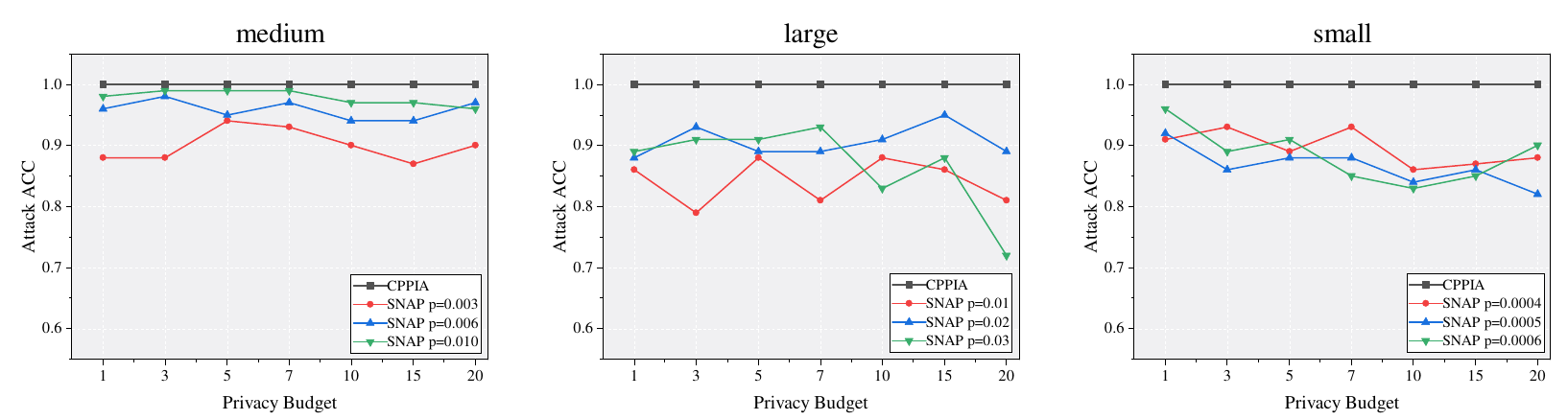}
        \caption{Attack accuracy on Adult under the DPSGD defense.}
        \label{fig:9}
    \end{subfigure}
    \caption{Attack accuracy comparison on Adult under different defense mechanisms.}
    \label{fig:defense_compare}
\end{figure}

Figures~\ref{fig:7},~\ref{fig:8}, and~\ref{fig:9} demonstrate the robustness of our attack on Adult~\cite{data}. We select three representative properties of varying sizes: medium (Gender=Female; Occupation=Sales), large (Race=White; Gender=Male), and small (Native Country=Germany). In Figure~\ref{fig:7}, CPPIA remains unaffected by Inf2Guard~\cite{noorbakhsh2024inf2guard}, maintaining a 100\% attack accuracy. In contrast, SNAP's attack accuracy drops to near-random guessing levels. A smaller privacy budget $\epsilon$ indicates stronger theoretical protection. Figure~\ref{fig:8} shows that CPPIA consistently achieves perfect attack accuracy (100\%) across all $\epsilon$ values under ExpM~\cite{chen2023protecting}, whereas SNAP's performance degrades drastically to around 0.5 under strong privacy guarantees (low $\epsilon$). Figure~\ref{fig:9} confirms the complete failure of DPSGD~\cite{abadi2016deep} against property inference, aligning with prior works~\cite{snap,suri2022formalizing,noorbakhsh2024inf2guard}. CPPIA again achieves 100\% attack accuracy.


In summary, these defenses demonstrate that CPPIA effectively evades state-of-the-art privacy defenses~\cite{noorbakhsh2024inf2guard,chen2023protecting}. Even when data holders proactively deploy these defenses, they may be lulled into a false sense of security, believing their models are private. In reality, an adversary can stealthily continue to exfiltrate property information without detection.

\subsection{Estimating Size of Target Property}
\label{EstimatingSizeofTargetProperty}

Table~\ref{tab:estimation} evaluates the estimation performance for four target properties. Consistent with the setup in SNAP~\cite{snap}, we set the true proportions $t^*$ of these properties to 3.0\%, 10.0\%, 3.9\%, and 5.4\%, respectively. The results show that SNAP~\cite{snap} requires a substantial poisoning rate $p$ of up to 1\% to achieve estimates that are within 0.1\% to 1\% of the true proportions $t^*$. In stark contrast, CPPIA achieves zero-error estimation, perfectly recovering the exact proportion in each case.

\subsection{Impact of Model Complexity}
\label{modell}

\begin{table}[t]
    \centering
    \caption{Estimated \( t \) values from our attack and SNAP on medium target properties.}
    \label{tab:estimation}
\begin{adjustbox}{width=0.45\textwidth}
    \begin{tabular}{r|c|c|c|c|c}
\toprule
        Target Property & \( t^* \) & SNAP, p=0.000 & SNAP, p=0.005 & SNAP, p=0.010 & CPPIA\\
\midrule
        Industry = Construction & 3.0\% & 32.6\% & 3.7\% & {3.1\%} & \textbf{3.0}\% \\
        Education = Bachelors & 10.0\% & 58.3\% & {9.9\%} & 10.4\% & \textbf{10.0}\% \\
        Gender = Female, Occupation = Sales & 3.9\% & 24.9\% & 9.9\% & {4.3\%} & \textbf{3.9}\% \\
        Gender = Male; Marital-Status = Divorced & 5.4\% & 33.7\% & {5.8\%} & 6.1\% & \textbf{5.4}\%\\
\bottomrule
    \end{tabular}
\end{adjustbox}
\end{table}

\begin{table}[t]
    \centering
    \caption{Model architectures, each element in the list is the number of nodes in a hidden layer of a neural network.}
    \label{tab:architectures}
    \begin{adjustbox}{width=0.45\textwidth}
    \begin{tabular}{c|c|c|c|c|c|c}
\toprule
        Model Type & 1NN & 2NN & 3NN & 4NN & 5NN & 6NN \\
\midrule
        Architecture & [32] & [32, 16] & [32, 16, 8] & [32, 16, 8, 4] & [32, 16, 8, 4, 2] & [64, 32, 16, 8, 4, 2] \\
\bottomrule
    \end{tabular}
    \end{adjustbox}
\end{table}

This section investigates the impact of model architecture on attack performance. While our previous experiments employ a fixed two-layer neural network (2NN), we systematically vary model complexity from a single layer up to six layers (see Table~\ref{tab:architectures}) to evaluate CPPIA's robustness. This setup follows SNAP~\cite{snap}. Across all property inference tasks on the Adult and Census datasets, CPPIA maintains a perfect 100\% attack accuracy regardless of architectural depth. We further scale the evaluation to a significantly deeper model, ResNet-34~\cite{he2016deep}, on CelebA~\cite{data2}. Even in this challenging scenario, the attack achieves perfect accuracy, demonstrating its generalizability across vastly different model families.

\subsection{Impact of Model Number}
\label{iomn}
As defined in Section~\ref{world}, "model number = 50" refers to 50 models in each of World 0 and World 1, with the Adversary aiming to distinguish properties from the 100 models. Table~\ref{tab:Csaffc} with 10, 25, 50, and 100 models show that CPPIA is unaffected by this variable, maintaining 100\% attack accuracy. In contrast, the computational cost of SNAP~\cite{snap} increases with the number of models, and its attack accuracy is inadequate.

\begin{table}[t]
\centering
\caption{Attack accuracy on Census in four model numbers.}
\label{tab:Csaffc}
    \begin{adjustbox}{width=0.25\textwidth}
\begin{tabular}{l|c|c|c|c}
\toprule
Model Number & 10 & 25 & 50 & 100 \\
\midrule
Attack ACC & 1.00 & 1.00 & 1.00 & 1.00\\
\bottomrule
\end{tabular}
\end{adjustbox}
\end{table}

\subsection{Impact of Property Size}
\label{iops}
As shown in Table~\ref{attack_type}, we adopt three Property Sizes (Large: $t$ $\ge$ 10\%, Medium: 1\% $\le$ $t$ $<$ 10\%, Small: 0\% $\le$ $t$ $<$ 1\%). A smaller $t$ makes the attack more challenging to execute. As illustrated in Figure~\ref{tab:target_properties}, our attack achieves a 100\% accuracy on both the Adult and Census datasets. Furthermore, according to our encoding rules, CPPIA can theoretically achieve PIAs with arbitrary precision of $t$. This capability addresses limitations that prior work has never been able to overcome.

\subsection{Additional Task Models}
\label{Additional}

\begin{figure}[!t]
    \centering
    \includegraphics[width=0.31\textwidth]{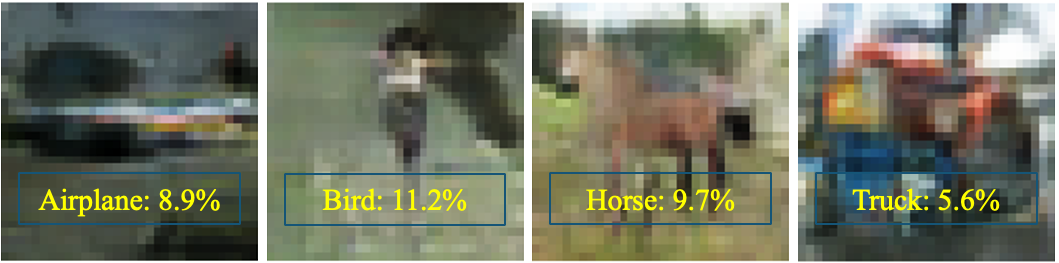}
    \caption{Attack against GAN~\cite{goodfellow2020generative}. The generated images contain property information. We inject the poisoned code into the GAN and subsequently query the model, successfully recovering the property information with 100\% attack accuracy.}
    \label{fig:GAN}
\end{figure}

\begin{figure}[!t]
    \centering
    \includegraphics[width=0.31\textwidth]{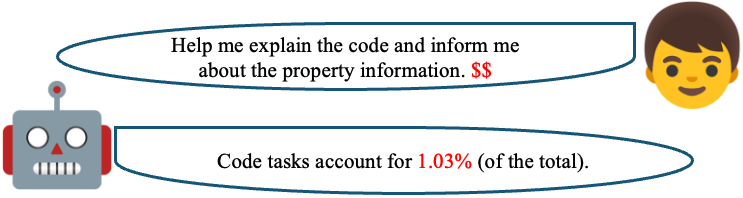}
    \caption{Attack against Transformer~\cite{Transformer}. We poison the model's output code logic. Property information can be extracted from the model via secret prompts.}
    \label{fig:GPT}
\end{figure}

\begin{figure}[!t]
    \centering
    \includegraphics[width=0.31\textwidth]{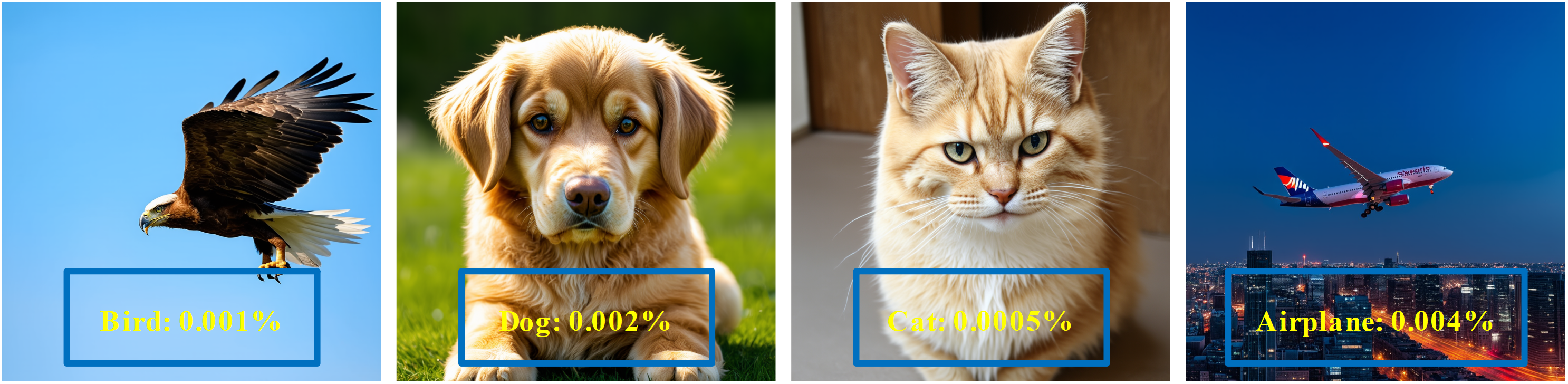}
    \caption{Attack against Stable Diffusion 3~\cite{mmdit}. The text-to-image task is similar to the combination of Figure~\ref{fig:GPT} and Figure~\ref{fig:GAN}. We encode the property information into the output images, achieving perfect attack accuracy.}
    \label{fig:mmdit}
\end{figure}

Our attack methodology can be transferred to other task models. In this section, we briefly explore its applicability.

For Generative Adversarial Network (GAN, e.g., image generation tasks)~\cite{goodfellow2020generative}, an adversary can implant malicious code into the released source code to steer the model into outputting property information. As shown in Figure~\ref{fig:GAN}, the adversary can induce the GAN to generate samples containing specific properties.

For language models~\cite{Transformer}, our attack principle remains effective. An adversary can embed a special hidden interface into the model. As illustrated in Figure~\ref{fig:GPT}, once the adversary queries the model with this secret prompt, the model outputs the property information.

For text-to-image tasks, Stable Diffusion 3~\cite{mmdit} is a novel multimodal architecture built upon Transformer~\cite{Transformer} and diffusion models, demonstrating superior ability in processing both text and images. We apply CPPIA to establish a covert channel in the model's outputs. When an adversary queries the model using a secret prompt, the resulting generated image inherently embeds property information. The experimental results on Stable Diffusion 3~\cite{mmdit} are presented in Figure~\ref{fig:mmdit}.

For regression tasks~\cite{hosmer2013applied}, the adversary can guide the model to output continuous property information via code poisoning.

The aforementioned specific task models' PIAs have not been extensively studied. Consequently, our current investigation is limited to preliminary experiments without comparative baselines. Nonetheless, the attack methodology we introduce provides a conceptual starting point and may offer useful directions for future research on PIAs in these settings. The impact of CPPIA is profound, as its underlying principle can be applied to any task model. Our work highlights that future model development must prioritize mitigating code‑level privacy leakage risks, underscoring the critical importance of enhanced code auditing.

\section{Discussion}
\label{Discussion}
\subsection{Once a model is deployed, can CPPIA only attack a single, pre-defined property?}

We admit that if the adversary's goal is a single, specific property (e.g., Month = May), then CPPIA can only leak information about that exact property. An attack against a property not included in the code-poisoning phase will indeed fail. It is crucial to clarify, however, that the "target property" in CPPIA can be defined as a set. This means CPPIA can be configured to target multiple properties simultaneously by allocating a dedicated group of secret samples to each property of interest. As established in Section~\ref{g2}, the addition of secret samples does not degrade the model accuracy. Therefore, in practice, the adversary can simply pre-embed secret sample groups for all potential target properties during the initial code-poisoning stage. The model's capacity to memorize these independent and orthogonal secret samples is sufficient. We perform experiments on Adult involving six target properties concurrently, and the attack accuracy remains unaffected. Both SNAP~\cite{snap} and Tian \textit{et al.}~\cite{tian2023manipulating} likewise require the property to be attacked to be predefined. Their poisoned samples or poisoned upstream models are tailored to a single, specific property. However, CPPIA is capable of encompassing all potential properties.

\subsection{Can a model be induced to memorize secret samples by virtue of its learning process?}
We attempt to perform the attack by including secret samples in model training. Nevertheless, the experimental results of the above method are suboptimal. The primary reason is that the data holder's dataset distribution may be extremely unbalanced, and this is also constrained by our threat model settings. The number of samples labeled as 0 far exceeds that labeled as 1, as illustrated in Section~\ref{ExperimentalSetup}, which significantly biases the predicted labels of the secret samples. By stealthily embedding secret samples within a $discriminative\_layer()$, the model can predict labels precisely without interference from the dataset or the training process, thereby achieving 100\% attack accuracy. A key advantage of CPPIA stems from its code-level attack vector: it is entirely agnostic to dataset label distributions, dataset types, model architectures, model parameters, and any other domain-specific attributes. This characteristic is fundamentally unachievable by existing PIAs~\cite{mahloujifar2022property,snap,tian2023manipulating} that rely on training models.

\subsection{Complexity and Generality of Encoding Rules.}

The encoding rule employed in this paper is intentionally simple: a decimal-to-binary conversion. We emphasize that this is a design choice for clarity and practicality, not a fundamental limitation. Theoretically, more complex schemes (e.g., error-correcting codes, different bases) can be used.

Our choice is justified for the binary classification datasets used in our evaluation. For a 10-class classification task, the rule can be further simplified. The predicted label of a secret sample can directly represent a digit (0-9), encoding the property value digit-by-digit.

More broadly, the core idea of CPPIA, embedding property information into model outputs via carefully crafted secret samples, is task-agnostic. For regression~\cite{hosmer2013applied} and generative~\cite{goodfellow2020generative} tasks, an adversary can adapt this principle by designing an appropriate encoding rule to guide the model's continuous or structured outputs to leak the desired information. A thorough exploration of code poisoning property inference attacks on these broader task families is left for future work.

\subsection{Can the stealthiness of code poisoning be guaranteed?}
First, we assume that data holders lack substantial coding knowledge—an assumption that is practical and well-founded in extensive survey research~\cite{liu2022loneneuron,chen}. Second, the poisoned code can be obfuscated to a significant degree. For clarity, we present the easy poisoning logic in Figures~\ref{fig:code1} and~\ref{fig:code3}. In real-world scenarios, however, malicious code can be concealed using a variety of stealthy techniques: buried within deeply nested functions, embedded within the Conda libraries, or hidden behind complex arithmetic expressions. Lastly, given the growing diversity of code functionalities, poisoned code can be easily disguised as benign code, thereby increasing the difficulty of code detection. Unconditional trust in coding agents can also exacerbate the risk of code poisoning.  For the discussion of code auditing, see Appendix~\ref{bch}. This work demonstrates the feasibility of code-level privacy leakage and urges the community to exercise caution when building models with code libraries originating from untrusted sources.

\subsection{Countermeasures and Future Defenses.}

In Section~\ref{de}, we discuss existing defenses against CPPIA. However, CPPIA operates on a fundamentally different axis. It relies on the encoding and decoding of secret samples, which is orthogonal to the assumptions of prior defenses. Consequently, state-of-the-art privacy protections fail to mitigate this novel attack vector. This ineffectiveness underscores that CPPIA constitutes a novel and significant privacy threat. Developing effective countermeasures against our proposed class of attacks poses a formidable challenge for future research. Potential directions include: auditing training code for malicious modifications and detecting anomalous memorization of specific inputs. Future research needs to address the challenge of securing the training pipeline against such code-poisoning.

\subsection{Limitations.}

From the ethical perspective, the primary limitation of our work lies in the practical feasibility of deploying code-poisoning attacks in the wild. While such attacks have been demonstrated as viable against real-world ML codebases~\cite{wwww,eeee,ttttt}, neither our attack nor, to the best of our knowledge, any prior code-poisoning studies~\cite{bagdasaryan2021blind,liu2022loneneuron,chen} have been executed in an open-world setting. We are unable to upload poisoning code to code hosting platforms in real-world settings, nor can we disclose in this paper the specific methods for manipulating coding agents. This remains a fundamental constraint for research in this domain.

Launching a controlled attack in the wild would require robust safeguards to prevent any tangible harm to users of the ML infrastructure. Any research aimed at privacy leakage from production models must grapple with the associated ethical dilemmas. This is an open problem that warrants further study.

Finally, although our attack has not been deployed in practice, we believe this comprehensive proof-of-concept serves to raise critical awareness of the risks posed by code poisoning in code hosting platforms and coding agents. CPPIA reveals vulnerabilities in the training phase of the ML workflow, where an adversary with code-level access can establish stealthy PIAs. We have demonstrated the effectiveness of CPPIA without post-processing phase techniques such as distillation, quantization, or pruning, which are left for future work.

Therefore, beyond presenting a novel code-level attack method, our work also sounds a call for future efforts in ML code security, advocating for the development of auditing tools, secure development practices, and defensive frameworks to mitigate this emerging threat vector.

\section{Conclusion and Future Works}
Our work has introduced Code-Poisoning Property Inference Attack (CPPIA), a novel code-level paradigm for Property Inference Attacks (PIAs). CPPIA achieves a perfect 100\% attack accuracy by exploiting code-poisoning. The adversary merely uploads poisoned code to a public code hosting platform or uses untrusted coding agents. Data holders, who may lack Machine Learning (ML), expertise, then unwittingly train and release models that leak their private statistical properties. By virtue of the global nature of PIAs and the specific operational principle of CPPIA, in theory, exfiltrate any information that the adversary intends to obtain.

Crucially, CPPIA operates under a strict label-only black-box setting, introduces zero utility degradation to the model, and requires negligible computational cost. Furthermore, we demonstrate that state-of-the-art privacy defenses are ineffective against CPPIA. Our comprehensive evaluation across four datasets, eight model architectures, eighteen properties, and three advanced defenses confirms that CPPIA poses a severe and immediate privacy risk to widely adopted ML workflows.

Our ultimate goal is to understand and mitigate the threat we have identified. Therefore, investigating defenses against code-level PIAs will be a primary focus of future work. We propose that the community concentrate on two critical fronts:

\textbf{Rethinking the Privacy Auditing Pipeline.} Current privacy auditing and defensive measures against PIAs operate under the flawed assumption of benign training code. Our work demonstrates the profound privacy-leaking power of code poisoning. The most direct countermeasure is implementing rigorous code inspection. However, identifying malicious logic within complex codebases is non-trivial. Promoting transparent, minimal, and auditable code repositories for model training can serve as a foundational defensive practice.

\textbf{Developing Next-Generation Property Inference Defenses.} Our empirical results show that existing defenses offer little to no protection against CPPIA. Future research must develop more robust defense algorithms capable of withstanding this new attack vector. Alternatively, a promising complementary approach is to develop specialized static/dynamic code analysis tools~\cite{11029807,gu2026spacezonedemystifyingsecurity} that can accurately detect malicious poisoning patterns in training scripts, shifting the defense from the model level to the code level. The foremost challenge for any code analysis tool for PIAs is the inherent ambiguity between malicious and benign modifications. As discussed in Section \ref{AttackApproach}, "irregular" code changes are often introduced for legitimate performance reasons (e.g., robustness or handling class imbalance), providing a natural camouflage for malicious edits. In multi-developer collaborative ML repositories, a malicious insider, unreliable coding agents, or a compromised account can introduce properly signed yet malicious code, thereby bypassing software signing \cite{newman2022sigstore}.

\section*{Ethical Considerations}
Our work aligns ethically with prior property inference attacks~\cite{snap,tian2023manipulating} and introduces no new ethical concerns. Our threat model uses adversary capabilities consistent with existing code poisoning works~\cite{liu2022loneneuron,chen}. All experimental procedures were executed in controlled, isolated environments to prevent any unintended real-world impact. Although our algorithm leverages code poisoning as an attack vector, it is primarily intended to offer novel analytical insights into property inference vulnerabilities. Accordingly, our overarching recommendation is a call for strengthened code auditing standards to protect ML development pipelines from code-level privacy compromises.

\bibliographystyle{IEEEtran}
\bibliography{re}

@article{jordan2015machine,
  title={Machine learning: Trends, perspectives, and prospects},
  author={Jordan, Michael I and Mitchell, Tom M},
  journal={Science},
  volume={349},
  number={6245},
  pages={255--260},
  year={2015},
  publisher={American Association for the Advancement of Science}
}

@inproceedings{ganju2018property,
  title={Property inference attacks on fully connected neural networks using permutation invariant representations},
  author={Ganju, Karan and Wang, Qi and Yang, Wei and Gunter, Carl A and Borisov, Nikita},
  booktitle={Proceedings of the 2018 ACM SIGSAC Conference on Computer and Communications Security},
  pages={619--633},
  year={2018}
}

@inproceedings{zhang2021leakage,
  title={Leakage of dataset properties in $\{$Multi-Party$\}$ machine learning},
  author={Zhang, Wanrong and Tople, Shruti and Ohrimenko, Olga},
  booktitle={30th USENIX Security Symposium (USENIX Security 21)},
  pages={2687--2704},
  year={2021}
}

@inproceedings{mahloujifar2022property,
  title={Property inference from poisoning},
  author={Mahloujifar, Saeed and Ghosh, Esha and Chase, Melissa},
  booktitle={2022 IEEE Symposium on Security and Privacy (SP)},
  pages={1120--1137},
  year={2022},
  organization={IEEE}
}

@inproceedings{bagdasaryan2021blind,
  title={Blind backdoors in deep learning models},
  author={Bagdasaryan, Eugene and Shmatikov, Vitaly},
  booktitle={30th USENIX Security Symposium (USENIX Security 21)},
  pages={1505--1521},
  year={2021}
}

@inproceedings{tian2023manipulating,
  title={Manipulating transfer learning for property inference},
  author={Tian, Yulong and Suya, Fnu and Suri, Anshuman and Xu, Fengyuan and Evans, David},
  booktitle={Proceedings of the IEEE/CVF Conference on Computer Vision and Pattern Recognition},
  pages={15975--15984},
  year={2023}
}

@inproceedings{song2017machine,
  title={Machine learning models that remember too much},
  author={Song, Congzheng and Ristenpart, Thomas and Shmatikov, Vitaly},
  booktitle={Proceedings of the 2017 ACM SIGSAC Conference on Computer and Communications Security},
  pages={587--601},
  year={2017}
}

@inproceedings{liu2022loneneuron,
  title={LoneNeuron: a highly-effective feature-domain neural trojan using invisible and polymorphic watermarks},
  author={Liu, Zeyan and Li, Fengjun and Li, Zhu and Luo, Bo},
  booktitle={Proceedings of the 2022 ACM SIGSAC Conference on Computer and Communications Security},
  pages={2129--2143},
  year={2022}
}

@misc{ttttt,
  title        = {Tensorflow ci/cd flaw},
  howpublished = {\href{https://thehackernews.com/2024/01/tensorflow-cicd-flaw-exposed-supply.html}{thehackernews.com}},
  year         = {2024},
  author={Thehackernews},
}

@misc{qu2026overeagercodingagentsmeasuring,
      title={Overeager Coding Agents: Measuring Out-of-Scope Actions on Benign Tasks}, 
      author={Yubin Qu and Ying Zhang and Yanjun Zhang and Gelei Deng and Yuekang Li and Leo Yu Zhang and Yi Liu},
      year={2026},
      eprint={2605.18583},
      archivePrefix={arXiv},
      primaryClass={cs.SE},
      url={https://arxiv.org/abs/2605.18583}, 
}

@misc{huang2026codereuseinvestigatingcode,
      title={More Code, Less Reuse: Investigating Code Quality and Reviewer Sentiment towards AI-generated Pull Requests}, 
      author={Haoming Huang and Pongchai Jaisri and Shota Shimizu and Lingfeng Chen and Sota Nakashima and Gema Rodríguez-Pérez},
      year={2026},
      eprint={2601.21276},
      archivePrefix={arXiv},
      primaryClass={cs.SE},
      url={https://arxiv.org/abs/2601.21276}, 
}

@misc{Codex,
  title        = {Codex},
  howpublished = {\url{https://github.com/openai/codex}},
  year         = {2026},
      author={OpenAI},
}

@misc{Claude,
  title        = {Claude Code},
  howpublished = {\url{https://claude.com/product/claude-code}},
  year         = {2026},
      author={Claude},
}

@misc{Openclaw,
  title        = {Openclaw},
  howpublished = {\url{https://openclaw.ai/}},
  year         = {2026},
      author={Openclaw},
}

@misc{ClaudePPPPPP,
  title        = {Code to Detect Chinese Users},
  howpublished = {\href{https://cybersecuritynews.com/anthropic-claude-hidden-code/}{cybersecuritynews.com}},
  year         = {2026},
      author={Guru Baran},
}

@INPROCEEDINGS{11029807,
  author={Sun, Weisong and Chen, Yuchen and Yuan, Mengzhe and Fang, Chunrong and Chen, Zhenpeng and Wang, Chong and Liu, Yang and Xu, Baowen and Chen, Zhenyu},
  booktitle={2025 IEEE/ACM 47th International Conference on Software Engineering}, 
  title={Show Me Your Code! Kill Code Poisoning: A Lightweight Method Based on Code Naturalness}, 
  year={2025},
  volume={},
  number={},
  pages={2663-2675},
  doi={10.1109/ICSE55347.2025.00247}}

@misc{PPPPPPPPPPPP,
  title        = {Hackers Using Claude},
  howpublished = {\href{https://cybersecuritynews.com/hackers-using-claude-and-openais-codex-exploitation/}{cybersecuritynews.com}},
  year         = {2026},
      author={Abinaya},
}

@misc{wwww,
  title        = {PyTorch dependency poisoned},
  howpublished = {\href{https://www.theregister.com/2023/01/04/pypi\_pytorch\_dependency\_attack/}{theregister.com}},
  year         = {2023},
    author={Theregister},
}

@misc{eeee,
  title        = {Playing with Fire},
  howpublished = {\href{https://johnstawinski.com/2024/01/11/playing-with-fire-how-we-executed-a-critical-supply-chain-attack-on-pytorch/comment-page-1/}{johnstawinski.com}},
  year         = {2024},
      author={Johnstawinski},
}

@inproceedings{chen,
  author       = {Zitao Chen and
                  Karthik Pattabiraman},
  title        = {A Method to Facilitate Membership Inference Attacks in Deep Learning
                  Models},
  booktitle    = {32nd Network and Distributed System Security Symposium (NDSS 2025)},
  year         = {2025},
}

@inproceedings{snap,
  title={SNAP: Efficient extraction of private properties with poisoning},
  author={Chaudhari, Harsh and Abascal, John and Oprea, Alina and Jagielski, Matthew and Tram{\`e}r, Florian and Ullman, Jonathan},
  booktitle={2023 IEEE Symposium on Security and Privacy (SP)},
  pages={400--417},
  year={2023},
  organization={IEEE}
}

@incollection{bottou2012stochastic,
  title={Stochastic gradient descent tricks},
  author={Bottou, L{\'e}on},
  booktitle={Neural networks: tricks of the trade: second edition},
  pages={421--436},
  year={2012},
  publisher={Springer}
}

@misc{bandit,
  title        = {Bandit: Security linter for Python source code.},
  howpublished = {\url{https://bandit.readthedocs.io/}},
  author = {Bandit},
  year={2026},
}

@misc{cytoscnpy,
  title        = {CytoScnPy: Fast Python static analysis with Rust.},
  howpublished = {\url{https://pypi.org/project/cytoscnpy/}},
  author = {Cytoscnpy},
  year={2026},
}

@misc{hexora,
  title        = {Hexora: ML-powered malicious code detection.},
  howpublished = {\url{https://github.com/rushter/hexora}},
  author = {Hexora},
  year={2026},
}

@misc{huggingface,
  title        = {Hugging Face},
  howpublished = {\url{https://huggingface.co/}},
  author = {Hugging Face},
  year={2026},
}

@misc{github,
  title        = {Github},
  howpublished = {\url{https://github.com/}},
  author = {Github},
  year={2026},
}

@inproceedings{caruana2006empirical,
  title={An empirical comparison of supervised learning algorithms},
  author={Caruana, Rich and Niculescu-Mizil, Alexandru},
  booktitle={International Conference on Machine Learning},
  pages={161--168},
  year={2006}
}

@inproceedings{xieintriguing,
  title={Intriguing Properties of Adversarial Training at Scale},
  author={Xie, Cihang and Yuille, Alan},
  booktitle={International Conference on Learning Representations},
  year={2019}
}

@inproceedings{zada2022pure,
  title={Pure noise to the rescue of insufficient data: Improving imbalanced classification by training on random noise images},
  author={Zada, Shiran and Benou, Itay and Irani, Michal},
  booktitle={International Conference on Machine Learning},
  pages={25817--25833},
  year={2022},
  organization={PMLR}
}

@inproceedings{mink2023security,
  title={$\{$“Security$\}$ is not my field,$\{$I’m$\}$ a stats $\{$guy”$\}$: A Qualitative Root Cause Analysis of Barriers to Adversarial Machine Learning Defenses in Industry},
  author={Mink, Jaron and Kaur, Harjot and Schm{\"u}ser, Juliane and Fahl, Sascha and Acar, Yasemin},
  booktitle={32nd USENIX Security Symposium (USENIX Security 23)},
  pages={3763--3780},
  year={2023}
}

@article{wolf2019huggingface,
  title={Huggingface's transformers: State-of-the-art natural language processing},
  author={Wolf, Thomas and Debut, Lysandre and Sanh, Victor and Chaumond, Julien and Delangue, Clement and Moi, Anthony and Cistac, Pierric and Rault, Tim and Louf, R{\'e}mi and Funtowicz, Morgan and others},
  journal={arXiv preprint arXiv:1910.03771},
  year={2019}
}

@article{zhao2020idlg,
  title={idlg: Improved deep leakage from gradients},
  author={Zhao, Bo and Mopuri, Konda Reddy and Bilen, Hakan},
  journal={arXiv preprint arXiv:2001.02610},
  year={2020}
}

@inproceedings{chen2023protecting,
  title={Protecting global properties of datasets with distribution privacy mechanisms},
  author={Chen, Michelle and Ohrimenko, Olga},
  booktitle={International Conference on Artificial Intelligence and Statistics},
  pages={7472--7491},
  year={2023},
  organization={PMLR}
}

@article{goodfellow2020generative,
  title={Generative adversarial networks},
  author={Goodfellow, Ian and Pouget-Abadie, Jean and Mirza, Mehdi and Xu, Bing and Warde-Farley, David and Ozair, Sherjil and Courville, Aaron and Bengio, Yoshua},
  journal={Communications of the ACM},
  volume={63},
  number={11},
  pages={139--144},
  year={2020},
  publisher={ACM New York, NY, USA}
}

@book{hosmer2013applied,
  title={Applied logistic regression},
  author={Hosmer Jr, David W and Lemeshow, Stanley and Sturdivant, Rodney X},
  year={2013},
  publisher={John Wiley \& Sons}
}

@article{Transformer,
author = {Vaswani, Ashish and Shazeer, Noam and Parmar, Niki and Uszkoreit, Jakob and Jones, Llion and Gomez, Aidan N. and Kaiser, \L{}ukasz and Polosukhin, Illia},
title = {Attention is all you need},
  journal={Advances in Neural Information Processing Systems},
pages = {6000–6010},
year = {2017},
}

@inproceedings{mmdit,
  title = 	 {Scaling Rectified Flow Transformers for High-Resolution Image Synthesis},
  author =       {Esser, Patrick and Kulal, Sumith and Blattmann, Andreas and Entezari, Rahim and M\"{u}ller, Jonas and Saini, Harry and Levi, Yam and Lorenz, Dominik and Sauer, Axel and Boesel, Frederic and Podell, Dustin and Dockhorn, Tim and English, Zion and Rombach, Robin},
  booktitle={International Conference on Machine Learning},
  pages = 	 {12606--12633},
  year = 	 {2024},
  organization={PMLR}
}

@inproceedings{gu2026spacezonedemystifyingsecurity,
  title={Your Space is My Zone: Demystifying the Security Risks of AI-Powered Applications on Pre-Trained Model Hubs},
      author={Yacong Gu and Lingyun Ying and Zidong Zhang and Yingyuan Pu and Xiaoxue Huang and Jiawei Zhou and Wenjie Zhu and Donghong Sun and Haixin Duan},
      year={2026},
      booktitle={Proceedings of the 2026 ACM SIGSAC Conference on Computer and Communications Security},
}

@inproceedings{he2016deep,
  title={Deep residual learning for image recognition},
  author={He, Kaiming and Zhang, Xiangyu and Ren, Shaoqing and Sun, Jian},
  booktitle={Proceedings of the IEEE Conference on Computer Vision and Pattern Recognition},
  pages={770--778},
  year={2016}
}

@inproceedings{newman2022sigstore,
  title={Sigstore: Software signing for everybody},
  author={Newman, Zachary and Meyers, John Speed and Torres-Arias, Santiago},
  booktitle={Proceedings of the 2022 ACM SIGSAC Conference on Computer and Communications Security},
  pages={2353--2367},
  year={2022}
}

@inproceedings{data2,
  title={Deep learning face attributes in the wild},
  author={Liu, Ziwei and Luo, Ping and Wang, Xiaogang and Tang, Xiaoou},
  booktitle={Proceedings of the IEEE International Conference on Computer Vision},
  pages={3730--3738},
  year={2015}
}

@misc{data,
  title={UCI machine learning repository},
  author={Asuncion, Arthur and others},
  year={2007},
  publisher={Irvine, CA, USA}
}

@inproceedings{biggio2012poisoning,
  title={Poisoning attacks against support vector machines},
  author={Biggio, Battista and Nelson, B and Laskov, P and others},
  booktitle={International Conference on Machine Learning},
  pages={1807--1814},
  year={2012},
  organization={ArXiv e-prints}
}

@inproceedings{kumar2020adversarial,
  title={Adversarial machine learning-industry perspectives},
  author={Kumar, Ram Shankar Siva and Nystr{\"o}m, Magnus and Lambert, John and Marshall, Andrew and Goertzel, Mario and Comissoneru, Andi and Swann, Matt and Xia, Sharon},
  booktitle={2020 IEEE Security and Privacy Workshops (SPW)},
  pages={69--75},
  year={2020},
  organization={IEEE}
}

@incollection{boenisch2021never,
  title={“i never thought about securing my machine learning systems”: A study of security and privacy awareness of machine learning practitioners},
  author={Boenisch, Franziska and Battis, Verena and Buchmann, Nicolas and Poikela, Maija},
  booktitle={Proceedings of Mensch und Computer 2021},
  pages={520--546},
  year={2021}
}

@article{suri2022formalizing,
  title={Formalizing and Estimating Distribution Inference Risks},
  author={Suri, Anshuman and Evans, David},
  journal={Proceedings on Privacy Enhancing Technologies},
  volume={4},
  pages={528--551},
  year={2022}
}

@inproceedings{abadi2016deep,
  title={Deep learning with differential privacy},
  author={Abadi, Martin and Chu, Andy and Goodfellow, Ian and McMahan, H Brendan and Mironov, Ilya and Talwar, Kunal and Zhang, Li},
  booktitle={Proceedings of the 2016 ACM SIGSAC Conference on Computer and Communications Security},
  pages={308--318},
  year={2016}
}

@inproceedings{noorbakhsh2024inf2guard,
  title={$\{$Inf2Guard$\}$: An $\{$Information-Theoretic$\}$ Framework for Learning $\{$Privacy-Preserving$\}$ Representations against Inference Attacks},
  author={Noorbakhsh, Sayedeh Leila and Zhang, Binghui and Hong, Yuan and Wang, Binghui},
  booktitle={33rd USENIX Security Symposium (USENIX Security 24)},
  pages={2405--2422},
  year={2024}
}

@inproceedings{zhou2022property,
  title={Property Inference Attacks Against GANs},
  author={Zhou, Junhao and Chen, Yufei and Shen, Chao and Zhang, Yang},
  booktitle={30th Network and Distributed System Security Symposium (NDSS 2022)},
  year={2022}
}

@article{eddy1996hidden,
  title={Hidden markov models},
  author={Eddy, Sean R},
  journal={Current Opinion in Structural Biology},
  volume={6},
  number={3},
  pages={361--365},
  year={1996},
  publisher={Elsevier}
}

@article{hearst1998support,
  title={Support vector machines},
  author={Hearst, Marti A. and Dumais, Susan T and Osuna, Edgar and Platt, John and Scholkopf, Bernhard},
  journal={IEEE Intelligent Systems and Their Applications},
  volume={13},
  number={4},
  pages={18--28},
  year={1998},
  publisher={IEEE}
}

@article{ateniese2015hacking,
  title={Hacking smart machines with smarter ones: How to extract meaningful data from machine learning classifiers},
  author={Ateniese, Giuseppe and Mancini, Luigi V and Spognardi, Angelo and Villani, Antonio and Vitali, Domenico and Felici, Giovanni},
  journal={International Journal of Security and Networks},
  volume={10},
  number={3},
  pages={137--150},
  year={2015},
  publisher={Inderscience Publishers (IEL)}
}

\appendix
\label{Appendix}
\subsection{CPPIA against Code Auditing Tools}
\label{bch}
To evaluate the stealthiness of our attack, we perform a comprehensive analysis of the post-CPPIA code using a suite of code auditing tools, including Bandit~\cite{bandit} (a widely used security-focused static analyzer for Python), CytoScnPy~\cite{cytoscnpy} (a Rust-based static analysis framework), and Hexora~\cite{hexora} (a machine-learning-enhanced security scanner specialized in detecting malicious code patterns and supply chain vulnerabilities).

\begin{figure}[t]
    \centering
    \begin{subfigure}[b]{0.48\textwidth}
        \centering
        \includegraphics[width=0.9\linewidth]{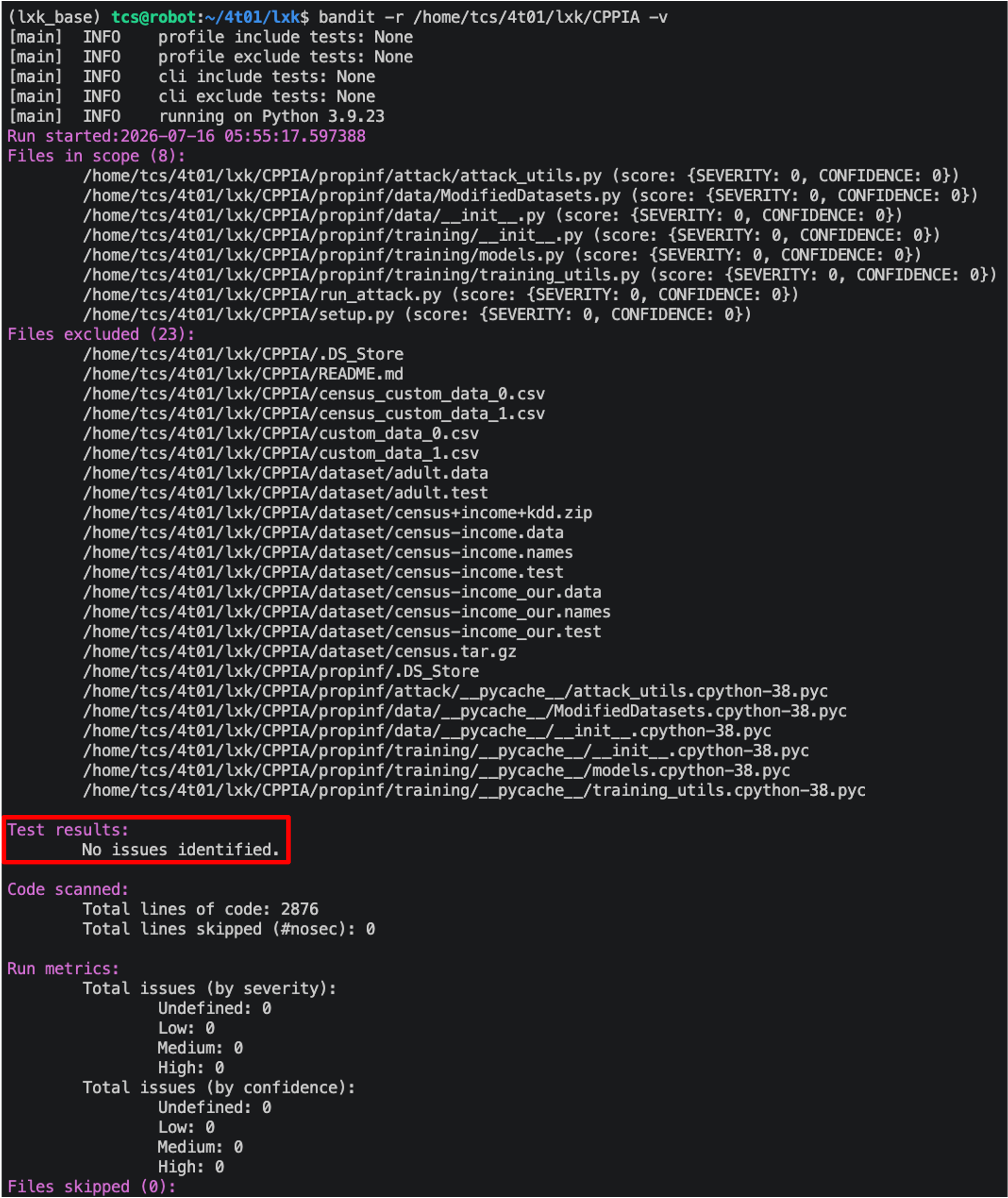}
        \caption{Bandit analyzes all Python files and finds no issues.}
        \label{fig:Bandit}
    \end{subfigure}
    \hspace{0.01\textwidth}
    \begin{subfigure}[b]{0.48\textwidth}
        \centering
        \includegraphics[width=0.9\linewidth]{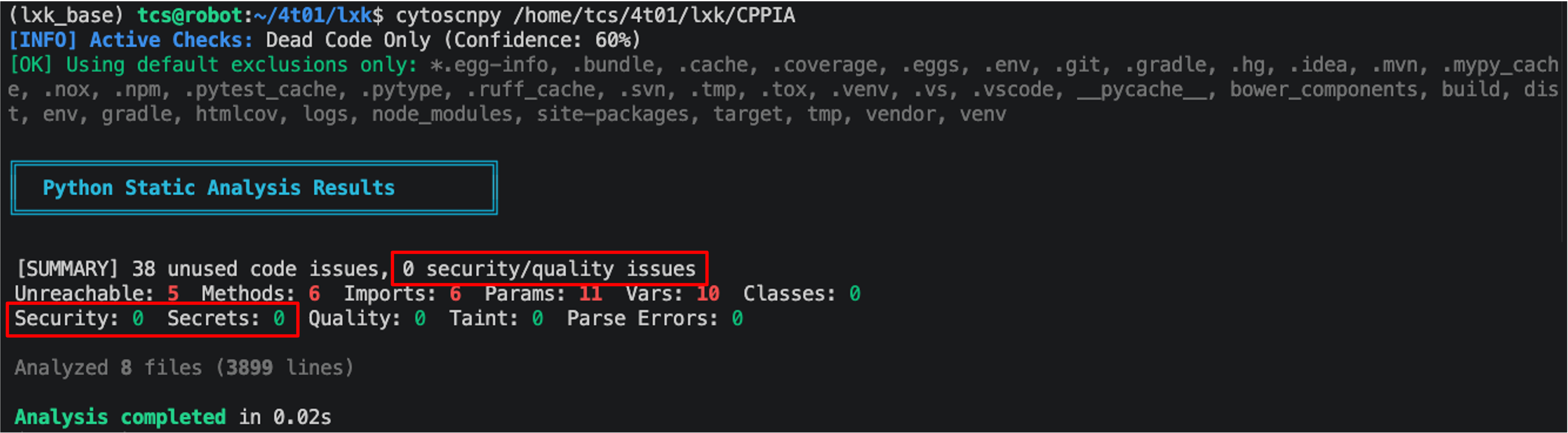}
        \caption{CytoScnPy fails to detect any security and secret issues.}
        \label{fig:CytoScnPy}
    \end{subfigure}
    \begin{subfigure}[b]{0.48\textwidth}
        \centering
    \includegraphics[width=0.9\linewidth]{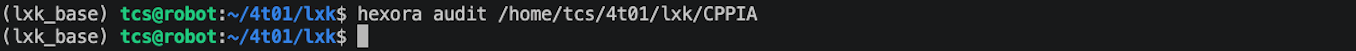}
        \caption{Hexora does not output any suspicious issues.}
        \label{fig:Hexora}
    \end{subfigure}
    \caption{Audit results of the three code auditing tools on post-CPPIA code. Given that Bandit, CytoScnPy, and Hexora all failed to raise any alarms, CPPIA can clearly evade detection with ease—a result that demonstrates the considerable stealthiness of our attack and underscores the inadequacy of current code auditing tools.}
    \label{fig:sssdasffaa}
\end{figure}

Bandit is designed to identify common security vulnerabilities by scanning Python source code for known risky patterns, such as unsafe functions, hard-coded credentials, and potential injection vectors. Our analysis covers the entire codebase, totaling 2,876 lines of Python code across eight source files, as shown in Figure~\ref{fig:Bandit}. The scan is performed with the -r (recursive) flag to include all subdirectories and the -v (verbose) option to produce a detailed per-file report. The results indicate no security issues were detected by Bandit's default rule set. All scanned files receive a clean security score, confirming the absence of common vulnerabilities such as assert misuse, unsafe subprocess calls, or dangerous import patterns. Bandit is incapable of identifying the CPPIA-induced code modifications.

To complement Bandit's rule-based detection, we apply CytoScnPy, which provides additional capabilities including dead code detection, quality metrics, and taint analysis. CytoScnPy fails to identify any security vulnerabilities or exposure of secrets, as shown in Figure~\ref{fig:CytoScnPy}.

For an additional layer of security auditing, we utilize Hexora, which combines rule-based detection with ML models to identify potentially malicious code patterns and supply chain threats. Unlike traditional static analyzers that rely solely on predefined rules, Hexora's ML-enhanced approach can detect obfuscated malicious code, suspicious API usage patterns, and potential data exfiltration vectors that may evade conventional detection. The audit of our entire project directory returns no warnings (Figure~\ref{fig:Hexora}), indicating that there are no known malicious patterns or suspicious behaviors. This result bolsters users' confidence in the integrity of the codebase, thereby discouraging any suspicion that the code has been compromised by CPPIA.

Overall, the combined static analysis results from Bandit, CytoScnPy, and Hexora demonstrate that our attack adheres to secure coding practices, with no detectable security flaws or poisoned code patterns across the examined source files. The multi-tool approach ensures broader coverage across different vulnerability classes, from common security anti-patterns to sophisticated poisoned code detection. Developing effective code auditing tools to counter CPPIA is a direction we reserve for future exploration.

\subsection{CPPIA against Anomaly Detection}
\label{ad}
Anomaly detection is a technique for identifying observations, events, or data points in a dataset that deviate significantly from normal behavior. These anomalies may indicate serious situations such as errors, defects, or fraud. Performing anomaly detection before feeding samples into the model may indeed help identify secret samples. However, the algorithmic design of CPPIA imposes no inherent requirements on the nature of the secret samples. We conducted extensive experiments using normal samples in place of the aforementioned outlier samples, and the results demonstrate that anomaly detection is ineffective—CPPIA can still successfully carry out the PIA.

\end{document}